\documentclass[]{spie}  
\usepackage[]{graphicx}

\title{Soft Gamma-ray Detector for the ASTRO-H Mission} 

\author{Hiroyasu~Tajima\supit{a}, 
Roger~Blandford\supit{a}, 
Teruaki~Enoto\supit{a},
Yasushi~Fukazawa\supit{b}, 
Kirk~Gilmore\supit{a},
Tuneyoshi~Kamae\supit{a},  
Jun~Kataoka\supit{c},
Madoka~Kawaharada\supit{d},
Motohide~Kokubun\supit{d},
Philippe~Laurent\supit{e},
Francois~Lebrun\supit{e},
Olivier~Limousin\supit{e},
Greg~Madejski\supit{a}, 
Kazuo~Makishima\supit{f},
Tsunefumi~Mizuno\supit{b},
Kazuhiro~Nakazawa\supit{f},
Masanori~Ohno\supit{d},
Masayuki~Ohta\supit{d},
Goro~Sato\supit{d},
Rie~Sato\supit{d},
Hiromitsu~Takahashi\supit{b}, 
Tadayuki~Takahashi\supit{d,f},
Takaaki~Tanaka\supit{a},
Makoto~Tashiro\supit{h},
Yukikatsu~Terada\supit{h},
Yasunobu~Uchiyama\supit{a},
Shin~Watanabe\supit{d,f},
Kazutaka~Yamaoka\supit{g} and
Daisuke~Yonetoku\supit{i}
\skiplinehalf
\supit{a} KIPAC, Stanford University, Stanford, CA 94305, USA\\
\supit{b} Department of Physical Science, Hiroshima University, Higashi-Hiroshima, Hiroshima 739-8526, Japan\\
\supit{c} Research Institute for Science and Engineering, Waseda University, Shinjuku-ku, Tokyo, 169-8050, Japan\\
\supit{d} Institute of Space and Astronautical Science, Sagamihara, Kanagawa 229-8510, Japan\\
\supit{e} IRFU / Service d'Astrophysique, CEA Saclay, 91191 Gif-sur-Yvette, Cedex France\\
\supit{f} Department of Physics, University of Tokyo, Bunkyo-ku, Tokyo 113-0033, Japan\\
\supit{g} Department of Physics, Aoyama-Gakuin University, Shibuya-ku, Tokyo 150-8366, Japan\\
\supit{h} Department of Physics, Saitama University, Saitama, Saitama 338-8570, Japan\\
\supit{i} Department of Physics, Kanazawa University, Kanazawa, Ishikawa 920-1192, Japan
}

\authorinfo{Further author information: (Send correspondence to H. Tajima)\\H. Tajima: E-mail: tajima@stanford.edu, Telephone: 1 650 926 3035}

\def\micron{{$\mu$m}}
\def\degC{{${}^\circ$C}}
\def\degree{{${}^\circ$}}
\def\degM{{{}^\circ}}
\def\cmsq{{cm${}^2$}}
\def\mmsq{{mm${}^2$}}

\newcommand{\lesssim}{\lower.5ex\hbox{$\buildrel < \over\sim$}}
\newcommand{\gtrsim}{\lower.5ex\hbox{$\buildrel > \over\sim$}}

\begin{document} 
\maketitle 

\begin{abstract}
ASTRO-H is the next generation JAXA X-ray satellite, intended to carry instruments with broad energy coverage and exquisite energy resolution.  The Soft Gamma-ray Detector (SGD) is one of ASTRO-H instruments and will feature wide energy band (40--600~keV) at a background level 10 times better than the current instruments on orbit.  
SGD is complimentary to ASTRO-H's Hard X-ray Imager covering the energy range of 5--80~keV.  The SGD achieves low background by combining a Compton camera scheme with a narrow field-of-view active shield where Compton kinematics is utilized to reject backgrounds.  
The Compton camera in the SGD is realized as a hybrid semiconductor detector system which consists of silicon and CdTe (cadmium telluride) sensors. Good energy resolution is afforded by semiconductor sensors, and it results in good background rejection capability due to better constraints on Compton kinematics.
Utilization of Compton kinematics also makes the SGD sensitive to the gamma-ray polarization, opening up a new window to study properties of gamma-ray emission processes.  
The ASTRO-H mission is approved by ISAS/JAXA to proceed to a detailed design phase with an expected launch in 2014.
In this paper, we present science drivers and concept of the SGD instrument followed by detailed description of the instrument and expected performance.
\end{abstract}


\section{INTRODUCTION}
\label{sec:intro}  
ASTRO-H, the new Japanese X-ray Astronomy Satellite\cite{NeXT08, Takahashi10} following the currently-operational Suzaku mission, aims to fulfill the following scientific goals:
\begin{itemize}
\item Revealing the large-scale structure of the universe and its evolution.
\item Understanding the extreme conditions of the universe.
\item Exploring the diverse phenomena of the non-thermal universe.
\item Elucidating dark matter and dark energy.
\end{itemize}
In order to fulfill the above objectives, the ASTRO-H mission hosts the following instruments: 
high energy-resolution soft X-ray spectrometer covering the 0.3--10~keV band, consisting 
of thin-foil X-ray optics (SXT, Soft X-ray Telescope) and a microcalorimeter array (SXS, 
Soft X-ray Spectrometer); soft X-ray imaging spectrometer sensitive over the 0.5--12~keV band, consisting of 
an SXT focussing X-rays onto CCD sensors (SXI, Soft X-ray Imager);  hard X-ray imaging spectrometer, 
sensitive over the 3--80~keV band, consisting of multi-layer-coated, focusing 
hard X-ray mirrors (HXT, Hard X-ray Telescope) and silicon (Si) and cadmium 
telluride (CdTe) cross-strip detectors (HXI, Hard X-ray 
Imager)\cite{Takahashi02-NeXT,Takahashi02,Takahashi03-SGD,Takahashi04-SGD,HXI08};  
and soft gamma-ray spectrometer covering the 40--600~keV band, utilizing the 
semiconductor Compton camera with narrow field of view (SGD, Soft Gamma-ray 
Detector)\cite{Takahashi02-NeXT,Takahashi02,Takahashi03-SGD,Takahashi04-SGD,Tajima05}.
The SXT-SXS and SGD systems will be developed by international collaboration led 
by the Japanese, US and European institutions.

The SXS will use a $6\times6$ element microcalorimeter array. 
The energy resolution is expected to be better than 7~eV. In conjunction
with the $\sim$6~m focal-length SXT, the field of view and the effective area will be, respectively, 
about 3 arc minutes and about 210~cm${^2}$.  The SXT-SXS system will provide 
accurate measurements of the temperature and the turbulence/macroscopic motions 
of intra-cluster medium in distant clusters of galaxies up to redshift of about 1, allowing studies 
of the formation history of the large scale structure of the universe, which 
in turn will eventually constrain the evolution of the dark energy.

The focal length of the HXT will be 12~m and the effective area will be larger 
than 200~cm${}^2$ at 50~keV.  The HXI detector utilizes four layers of double-sided Si 
strip detectors overlaid on a double-sided CdTe strip detector with a 
BGO (Bi${}_4$Ge${}_3$O${}_{12}$) active shield.  The extremely low background 
of the HXT-HXI system will improve the sensitivity in 20--80~keV range by almost 
two orders of magnitude as compared to conventional non-imaging detectors in this energy band.
The search for highly absorbed active galactic nuclei and understanding their 
evolution is one of main science topics of the HXT-HXI.

The SGD also utilizes semiconductor detectors using Si and CdTe pixel 
sensors with good energy resolution ($\lesssim$2~keV) for the Compton camera, which was made possible by recent progress on the development of high quality CdTe sensors\cite{Takahashi01b,Watanabe05,Watanabe09,Takeda09}.
The BGO active shield provides a low background environment by rejecting the
majority of external backgrounds.  Internal backgrounds are rejected based 
on the inconsistency between the constraint on the incident angle of 
gamma rays from Compton kinematics and that from the narrow FOV (field 
of view) of the collimator.  This additional background rejection by 
Compton kinematics will improve the sensitivity by an order of magnitude 
in the 40--600~keV band compared with the currently operating space-based instruments.
Science objectives of the SGD include studies of particle acceleration in 
various sources via measurement of non-thermal emission and the high-energy cutoff, 
origin of the emission in the GeV gamma-ray emission though the observation of non-thermal bremsstrahlung signatures expected in the SGD band, 
and searches for origin of 511~keV emission from electron-positron annihilation.
In addition, the SGD will be sensitive to the polarization in the 50--200~keV band from 
a number of accreting Galactic black hole and neutron star binaries, and 
for active galactic nuclei in flaring states.

The ASTRO-H mission has just completed the preliminary design 
phase, aimed at verifying that the design of the system/subsystems/components 
will meet the mission requirements with sufficient reliability to accomplish 
the mission objectives, and to assure that system/subsystems/components 
designs are feasible in terms of technology and schedule via design analysis, 
fabrication and tests of bread board models.  The expected launch date is in 2014.

\section{Science Requirements and Drivers}
The mission-level science objectives described above require the SGD to provide 
spectroscopy up to 600~keV for over 10 super-massive black holes with fluxes equivalent to 
1/1000 of the Crab Nebula (as measured over the 2--10~keV band, assuming the spectrum 
to be a power-law with spectral index of 1.7).  This mission-level science requirement 
defines the following instrument-level requirements for the SGD:  
\begin{itemize}
\item Effective area for the detector must be greater than 20~\cmsq\ at 100~keV to obtain sufficient number of photon in a reasonable observation time (typically 100~ks);  
\item Field of view must be 0.6\degree\ at 150~keV or less to minimize source confusion;
\item Energy resolution must be better than 2~keV to identify nuclear lines from activation backgrounds.
\end{itemize}
The SGD instrument with capabilities defined above will
determine the non-thermal emission processes for a large range
of celestial sources (via the measurement of 
broad-band spectral shape and and high-energy cutoff), with the goal 
of studying particle acceleration in GeV band.  With some sources, 
parameters of non-thermal bremsstrahlung 
processes will be determined;  and finally, SGD will enable the identification of the 
origin of 511~keV emission line, arising from electron-positron annihilation.

Measurements of spectra up to 600~keV for more than 10 AGNs (Active Galactic Nuclei) 
will enable a probe of existence of spectral breaks above 100~keV.  Measurements of such 
spectral breaks will play a crucial role in solving the question on the origin 
of the soft gamma-ray emissions in AGN (whether the emission arises 
from the accretion disk or relativistic electrons in the jet).  
The detailed spectral measurements are expected to contribute to 
understanding of the soft gamma-ray emission in more than 10 X-ray 
pulsars and magnetars. 

The SGD is also expected to be able to measure the spectrum 
of supernova remnants (with the prime example of Cas~A), to determine
whether it is indeed dominated by non-thermal bremsstrahlung, as
expected on theoretical grounds.  The soft gamma-ray flux measured by 
the SGD can determine the magnetic field of Cas~A by combining the SGD data 
with those from other wavelength:  this is essential when estimating the 
fluxes and spectra of electrons and protons accelerated in Cas~A.

Perhaps the most unique SGD parameter is that Compton kinematics utilized in the SGD 
yield good sensitivity to the polarization in the 50--200~keV band 
from several Galactic black holes and neutron stars, and some AGNs in flare states.
Detection of the gamma-ray polarization from these sources will bring new probes 
into the gamma-ray emission mechanism.  Moreover, the detection of X-ray / soft gamma-ray 
polarization from sources at the cosmological distance will place stringent constraints 
on the violation of Lorentz invariance, which has a profound impact on the fundamental physics.
Since X-ray polarization is largely unexplored, discovery potential is very high.

In summary, the SGD is expected to provide essential data towards studies 
of the origin of CXB (Cosmic X-ray background), particle acceleration in SNR, 
origin of the hard X-ray emission from the vicinity of accreting 
black holes such as X-ray binaries, 
the Galactic center, and AGN, and non-thermal emission from galaxy clusters.

\section{Instrument Concept}
\begin{figure}[bth]
   \centering
   \begin{tabular}{ll}
   \includegraphics[height=8cm]{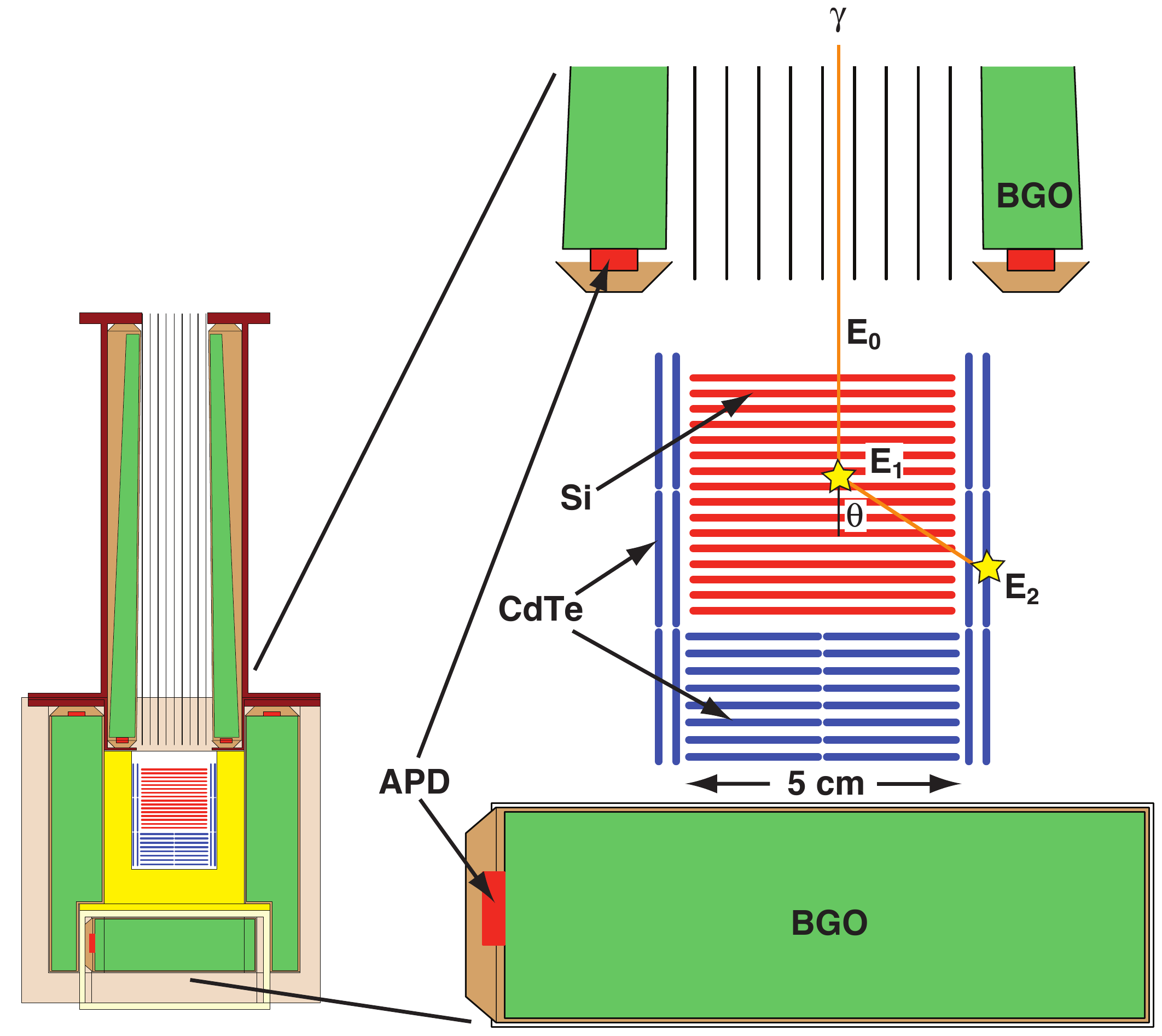} 
   \end{tabular}
   \caption{Conceptual drawing of an SGD Compton camera unit.}\label{fig:SGD-concept}
   \end{figure}

The SGD concept originates from Hard X-ray Detector (HXD)\cite{HXD} onboard Suzaku satellite.
The HXD consists of Si photodiodes and GSO scintillators with BGO active shield and copper passive collimator, and achieved the best sensitivities in the hard X-ray band.
The SGD replaces the Si photodiodes and GSO scintillators in HXD with the Compton camera, which provides additional background rejection capabilities based on Compton kinematics.
Figure~\ref{fig:SGD-concept} (a) shows a conceptual drawing of a SGD unit.
A BGO collimator defines $\sim$10\degree\ FOV of the telescope for high energy photons while a fine collimator restricts the FOV to $\lesssim$0.6\degree\ for low energy photons (\lesssim 150~keV), which is essential to minimize the CXB (cosmic X-ray backgrounds) and source confusions.
Scintillation light from the BGO crystals is detected by avalanche photo-diodes (APDs) allowing a compact design compared with phototubes.

The hybrid design of the Compton camera module incorporates both pixelated Si and CdTe detectors.
The Si sensors are used as the scatterer since Compton scattering is the dominant process in Si above $\sim$50~keV compared with $\sim$300~keV for CdTe.
The Si sensors also yield better constraints on the Compton kinematics because of smaller effect from the finite momentum of the Compton-scattering electrons (Doppler broadening) than CdTe (approximately by a factor of two).
The CdTe sensors are used as the absorber of the gamma ray following the Compton scattering in the Si sensors.
Combination of two materials with low and high $Z$ (atomic number) are also beneficial for lowering backgrounds since neutron scattering is suppressed in high-Z material and activation backgrounds are negligible in low-$Z$ material.
Note that neutron and activation backgrounds are the dominant background contributions in the SGD.

We require each SGD event to interact twice in the Compton camera, once by Compton scattering in a Si sensor, and then by photo-absorption in a CdTe sensor.
Once the locations and energies of the two interactions are measured, as shown in Figure~\ref{fig:SGD-concept}, the Compton kinematics can be calculated by the direction of the incident photon with the formula,
\begin{eqnarray}
\cos\theta &=& 1+\frac{m_ec^2}{E_2+E_1}-\frac{m_ec^2}{E_2},
\label{eq:kinematics}
\end{eqnarray}
where $\theta$ is the polar angle of the Compton scattering, and $E_1$ and $E_2$ are the energy deposited in each photon interaction.
The high energy resolution of the Si and CdTe devices is essential in reducing the uncertainty of $\theta$.
The angular resolution is limited to $\sim$8\degree\ at 100~keV due to the Doppler broadening and $\sim$3\degree\ at 600~keV due to pixel size of the semiconductor sensor.
We require that the incident photon angle inferred from the Compton kinematics is consistent with the FOV, which dramatically reduces dominant background sources such as radio-activation of the detector materials and neutrons.
Low background realized by the Compton kinematics is the key feature of SGD since the photon sensitivity of SGD is limited by the backgrounds, not the effective area.

As a natural consequence of the Compton approach used to decrease backgrounds, SGD is quite sensitive to X/gamma-ray polarization, thereby opening up a new window to study the geometry of the particle acceleration and emission regions, and the magnetic field in compact objects and astrophysical jets.
The Compton scattering cross section depends on the azimuth Compton scattering angle with respect to the incident polarization vector as;
\begin{eqnarray}
\frac{\delta\sigma}{\delta\Omega} \propto \left( \frac{E_\gamma'}{E_\gamma}\right)^2\left(\frac{E_\gamma'}{E_\gamma}+\frac{E_\gamma}{E_\gamma'}-2\sin^2\theta\cdot\cos\phi\right),
\label{eq:polarization}
\end{eqnarray}
where $\phi$ and $\theta$ are the azimuth and polar Compton scattering angles, and $E_\gamma$ and $E_\gamma'$ are incident and scattered photon energies.
It shows that the $\phi$ modulation is largest at $\theta=90^\circ$, \emph{i.e.} perpendicular to the incident polarization vector.

\section{Instrument Design}
\begin{figure}[bth]
   \centering
   \begin{tabular}{ll}
   (a) & (b) \\
   \includegraphics[height=6cm]{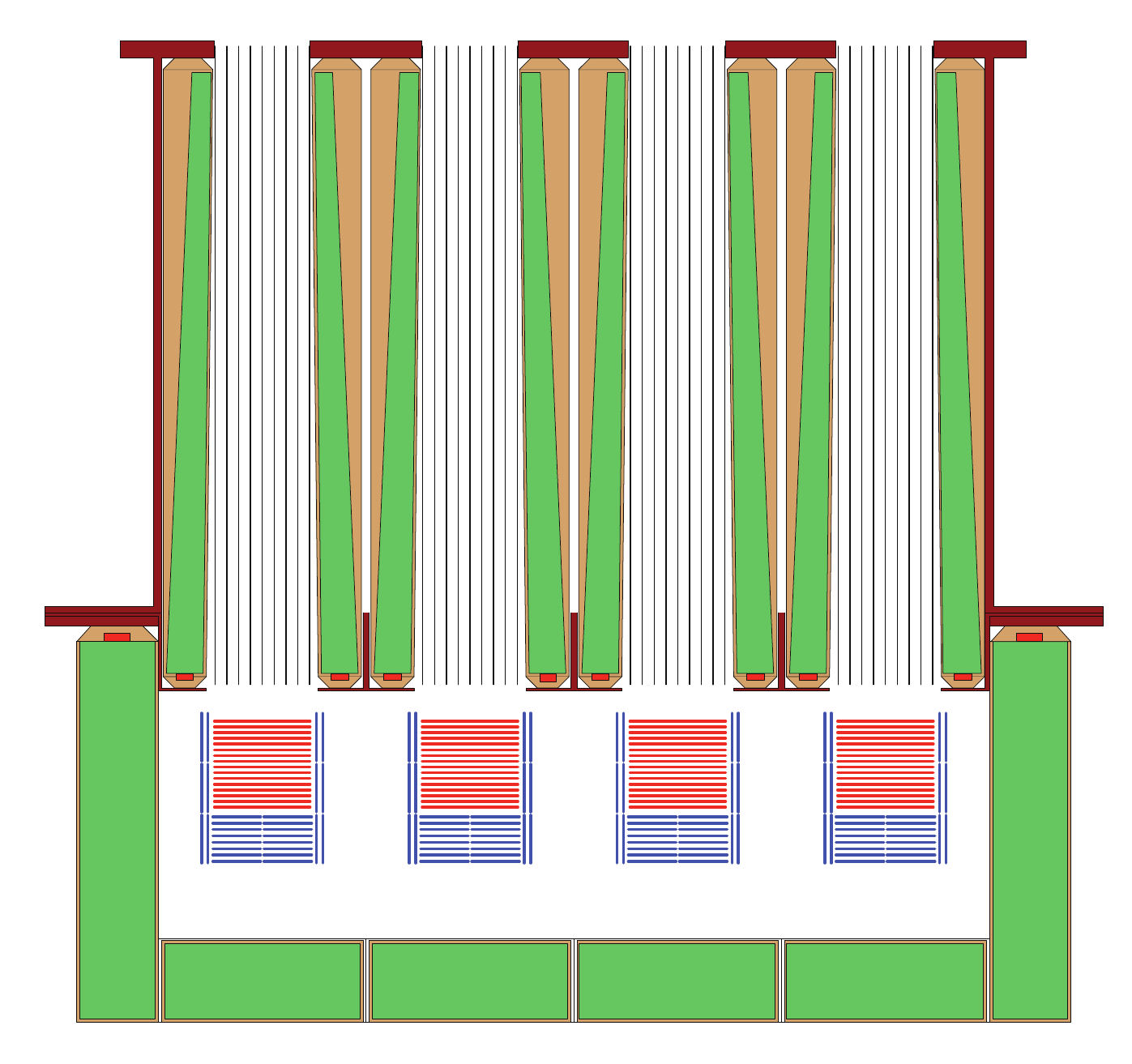} \hspace*{0.2cm} &
   \includegraphics[height=4.5cm]{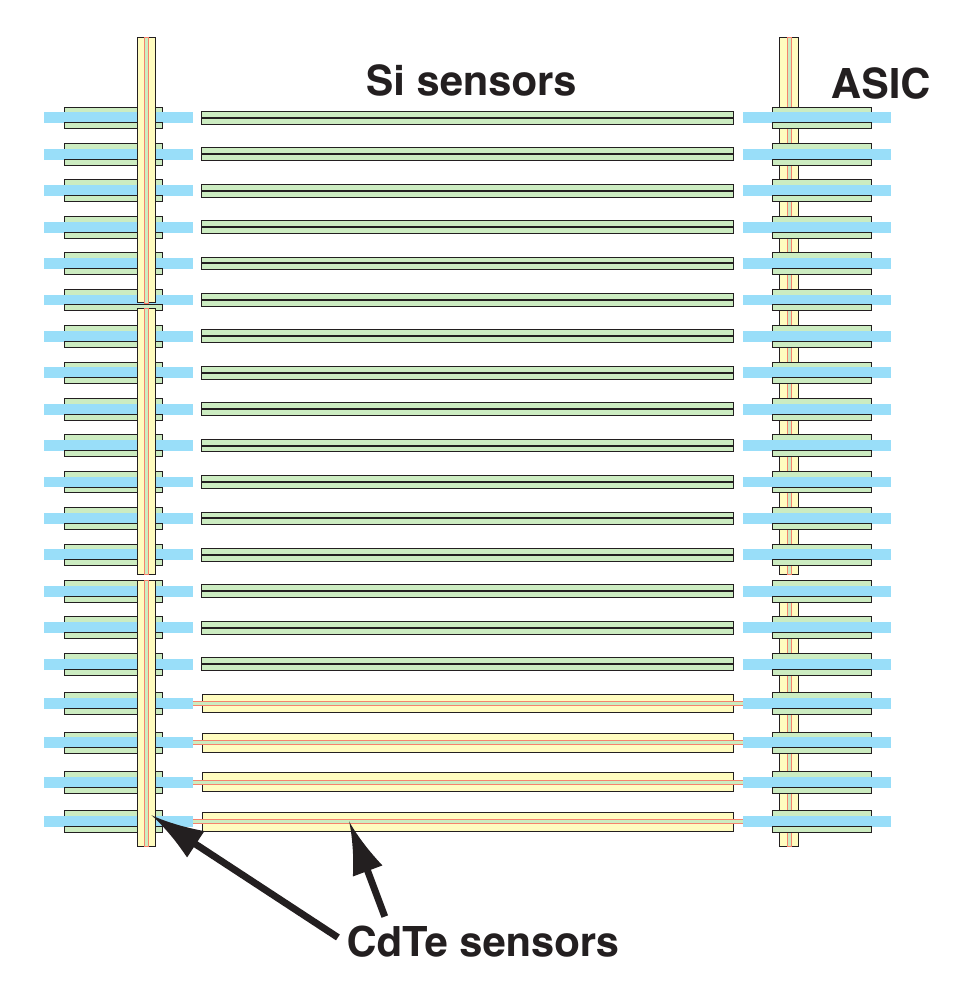} 
   \includegraphics[height=4.5cm]{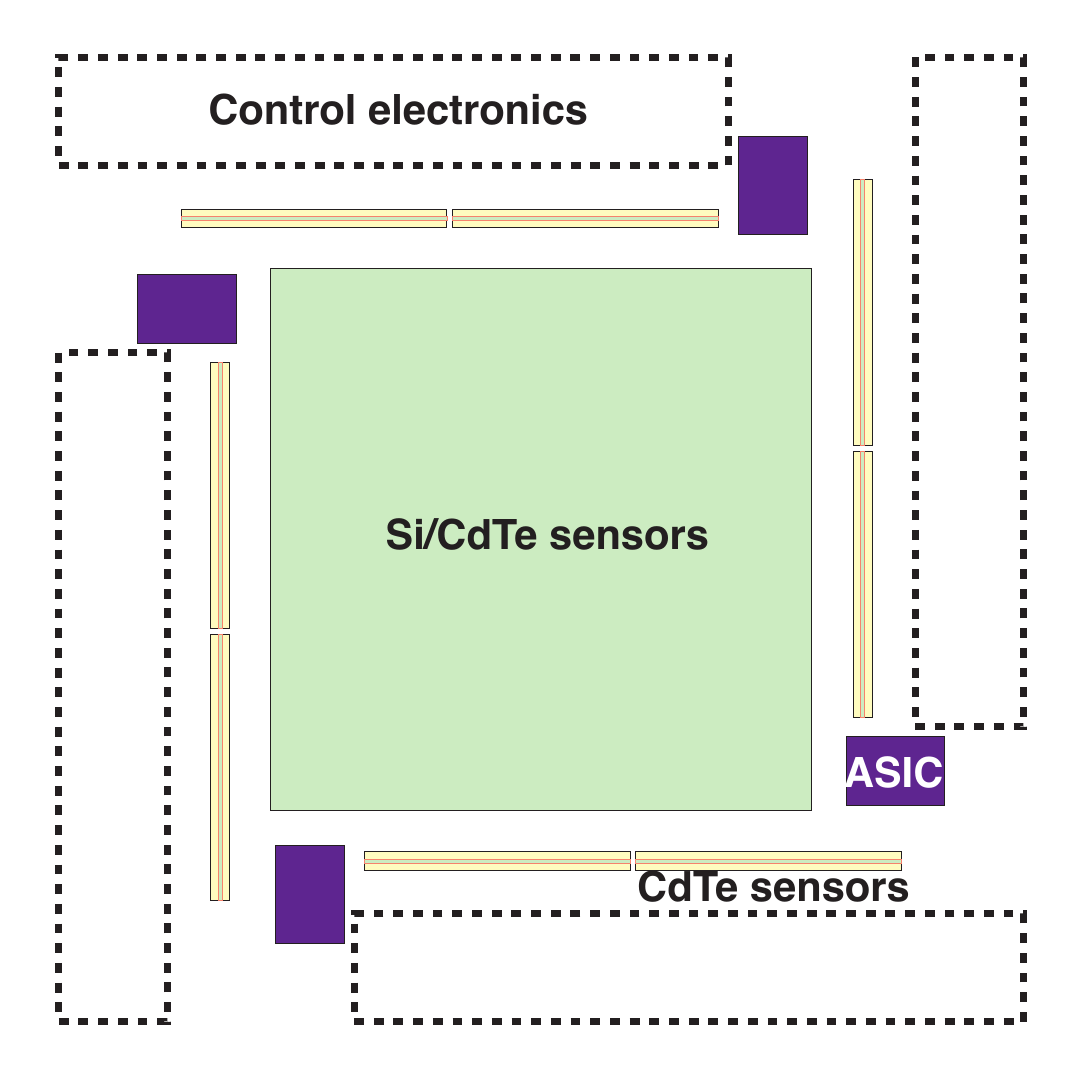} 
   \end{tabular}
   \caption{Schematic drawing of (a) an SGD-S and (b) sensor configuration of a Compton camera.}\label{fig:SGD-design}
   \end{figure}

SGD consists of two identical set of a SGD-S, two SGD-AE, a SGD-DPU and a SGD-DE.
SGD-S is a detector body that includes a $4\times 1$ array 
of identical Compton camera modules surround by BGO shield 
units and fine passive collimators as shown in Figure~\ref{fig:SGD-design} (a).
Two SGD-S are mounted on opposite sides of the spacecraft 
side panels to balance the weight load since it has a high mass (150~kg).
It was determined that a $2\times 2$ array arrangement is preferred as 
it allows an increase of the BGO thickness for the same weight and can 
also keep a symmetry against 90\degree\ rotation which is 
important for polarization measurements.  However, the current 
$4\times 1$ is employed to minimize the deformation of the 
spacecraft side panel.  SGD cooling system is attached to the cold 
plate of the SGD-S housing.  APD CSA (charge-sensitive amplifier) 
box and HV (high voltage) power supply are also attached to the SGD housing.
SGD-AE is an electronics box that provides power management and housekeeping (HK) functions for Compton camera system and APD readout system.
It also performs APD signal processing.  SGD-DPU functions as 
a digital interface to the SGD-DE via SpaceWire network standard 
and also houses SGD-PSU (Power Supply Unit) inside.
SGD-DE includes a microprocessor and performs data processing for 
event and HK data, and is connected to the satellite SpaceWire network.
Topology of the SpaceWire network is designed to be redundant.
Data can be routed via another DPU if one of DPU-DE or DE-router 
connections is broken.  In addition, a spare DE shared by all instruments 
on the satellite is included in the payload: the data can be routed to the spare DE if one of the SGD-DEs malfunctions.
Design details of each component are described below.

\subsection{Compton camera}
The Compton camera consists of 32 layers of Si sensors 
and 8 layers of CdTe sensors surrounded by 2 layers of 
CdTe sensors as shown in Figure~\ref{fig:SGD-design} (b).
The location of the CdTe sensors on the side is slightly 
displaced in the horizontal direction to allow placement 
of readout ASIC (Application Specific Integrated Circuit) at the corner of the sensor.  
This arrangement allows a placement of the CdTe sensor 
on the side very close to the stacked Si and CdTe sensors 
to maximize the coverage of the photons scattered by the Si sensors.
In addition to sensor modules, the Compton camera holds 
an ACB (ASIC controller board) and four ADBs (ASIC driver boards).
The ACB holds an FPGA (filed programmable gate array) 
that controls the ASIC and communicates with SpaceWire 
interface by a serial link.  The ADB buffers control signals 
from the ACB and sends control signals to 52 ASICs, and also 
provides a current limiter to power the ASICs.

The mechanical structure of the Compton camera needs to hold all components described above within the size of $11\times11\times12$~cm$^3$. 
This size constraint is imposed to minimize the size 
of BGO active shield since the BGO is the dominant 
contribution to the total weight of the SGD-S.
Another important requirement 
for the mechanical structure is cooling of the sensors.
The temperature of all sensors needs to be maintained 
to within 5\degC\ of the cold plate interface at the bottom of the Compton camera.
\begin{figure}[htbp] 
   \centering
   \begin{tabular}{ll}
   (a) & (b) \\
   \includegraphics[height=5cm]{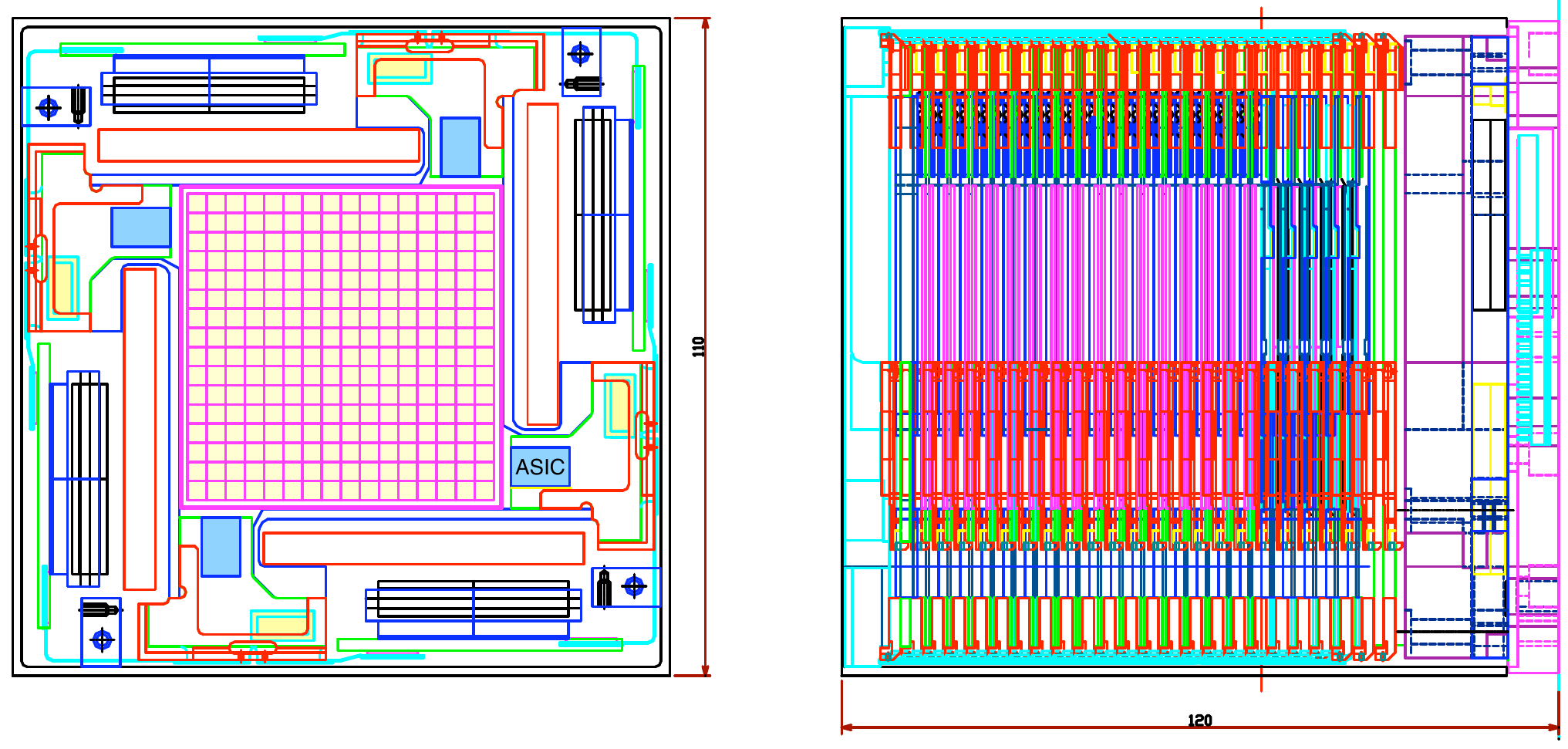} &
   \includegraphics[height=5cm]{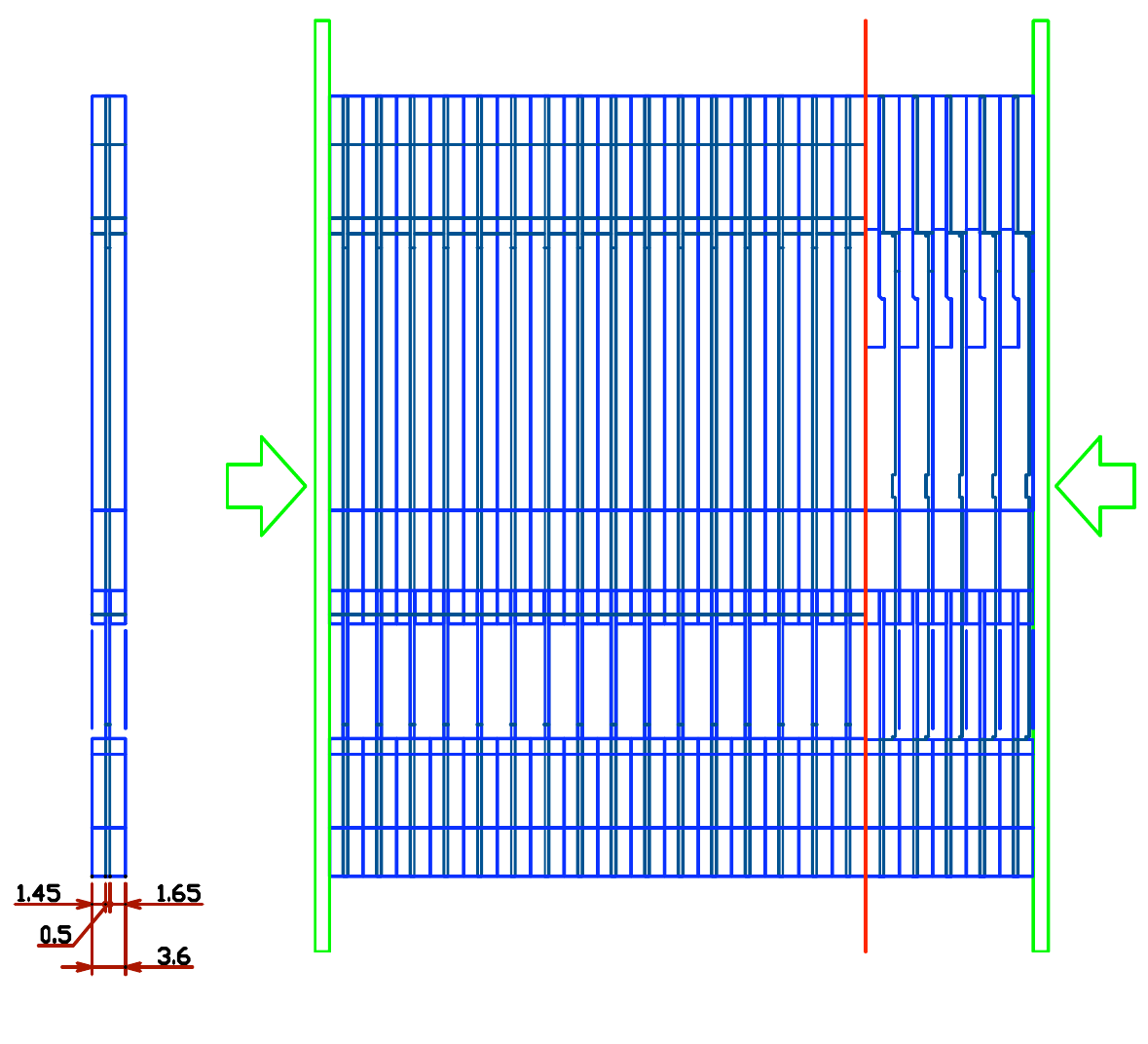} 
   \end{tabular}
   \caption{(a) Drawing of Compton camera structure. (b) Drawing of a stack of Si and CdTe sensor tray modules. Note they are facing left in the side view.}
   \label{fig:CC-structure}
\end{figure}
\begin{figure}[htbp] 
   \centering
   \begin{tabular}{ll}
   (a) & (b) \\
   \includegraphics[height=5cm]{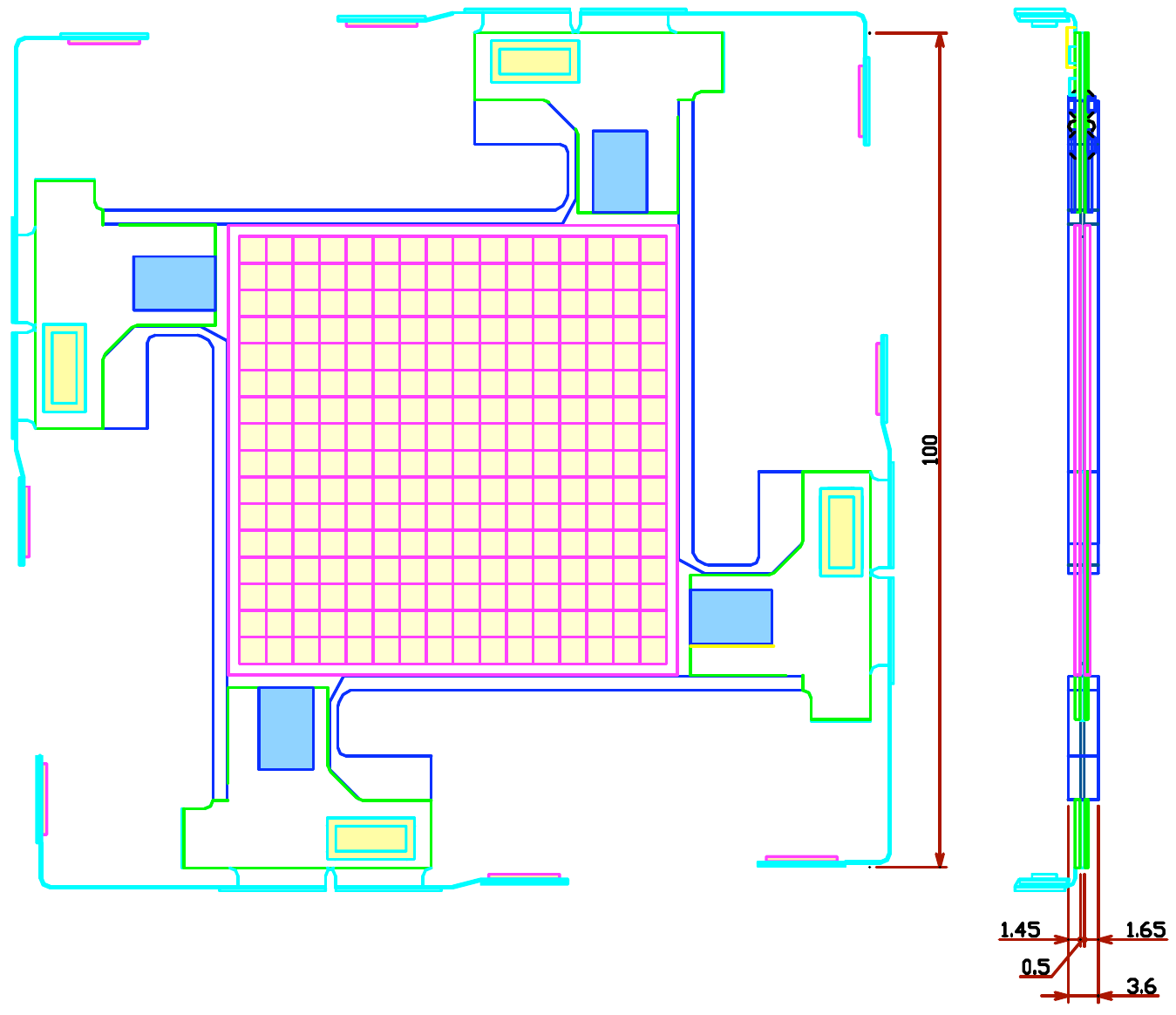} \hspace*{1cm} &
   \includegraphics[height=5cm]{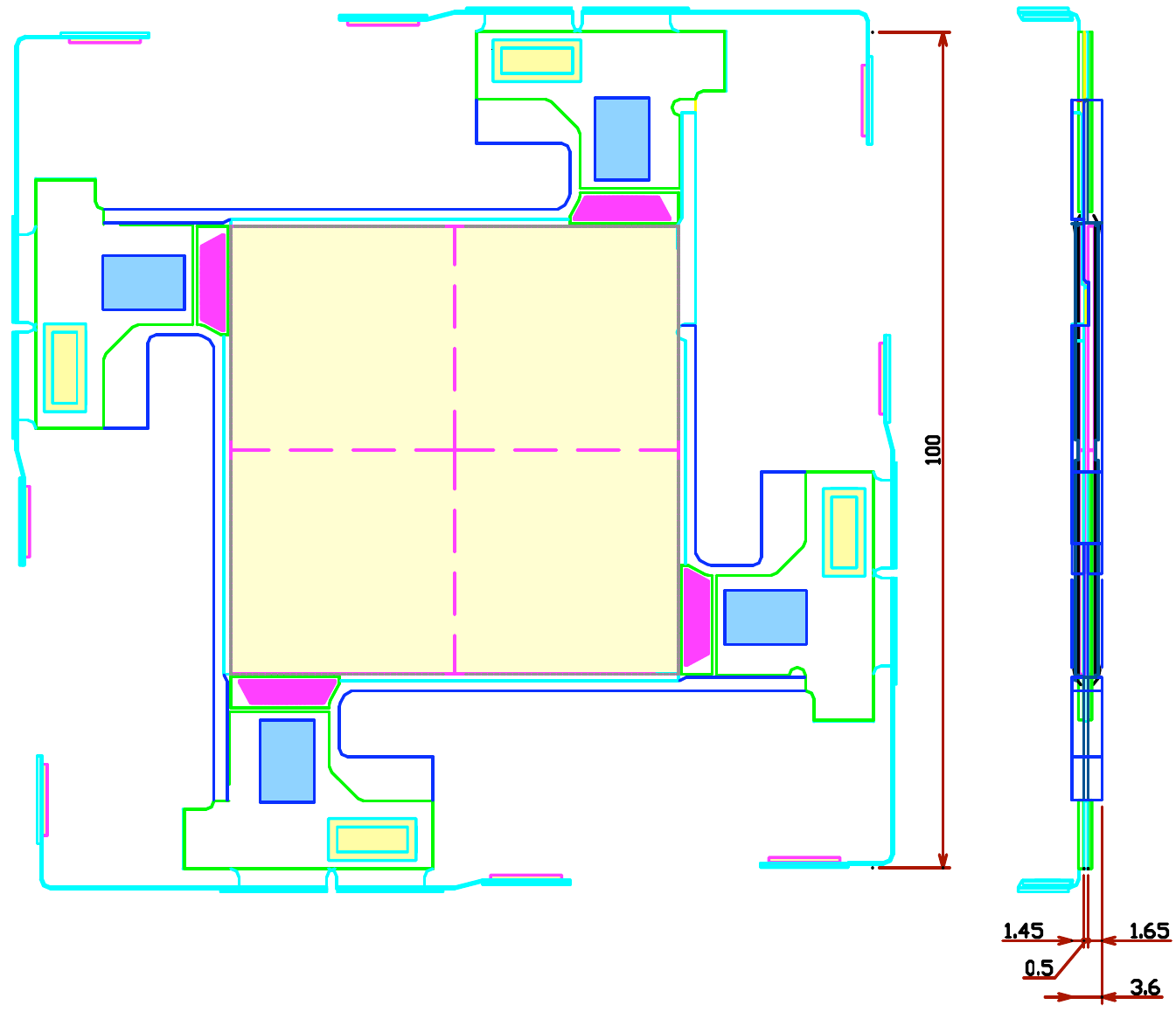} 
   \end{tabular}
   \caption{Top and side views of (a) Si and (b) CdTe sensor tray modules.}
   \label{fig:CC-trays}
\end{figure}

Figure~\ref{fig:CC-structure} (a) shows the mechanical support structure of the Compton camera.
The Compton Camera consists of a stack of Si and CdTe sensor trays as shown 
in Figure~\ref{fig:CC-structure} (b), four CdTe sensor modules on the side, 
and also top and bottom frames to hold them together.  The top and bottom 
frames are held together by four pillars with M3 screws.  Each ADB is 
attached to the side CdTe sensor module and an ACB is attached to the bottom frame.
The material of the camera structure must have the CTE (Coefficient 
for Thermal Expansion) close to those of Si and CdTe sensors (a few~\micron/m/K).
Currently, it is planned to employ PEEK (polyether ether ketone) loaded with 
carbon fibers for the trays and the top frame where a low-$Z$ material is required, and titanium for the pillars and the bottom frame.  
Since the carbon fiber-filled PEEK is conductive, 
trays need to be conformal-coated by Parylene (commercial name of Xylylen polymers) 
to avoid shorting of the bias voltage for the sensors.
Parylene can produce pinhole-free coating with high resistivity, 
uniform thickness and chemical tolerance.

\begin{figure}[htbp] 
   \centering
   \begin{tabular}{ll}
   (a) & (b)\\
   & \includegraphics[height=2.5cm]{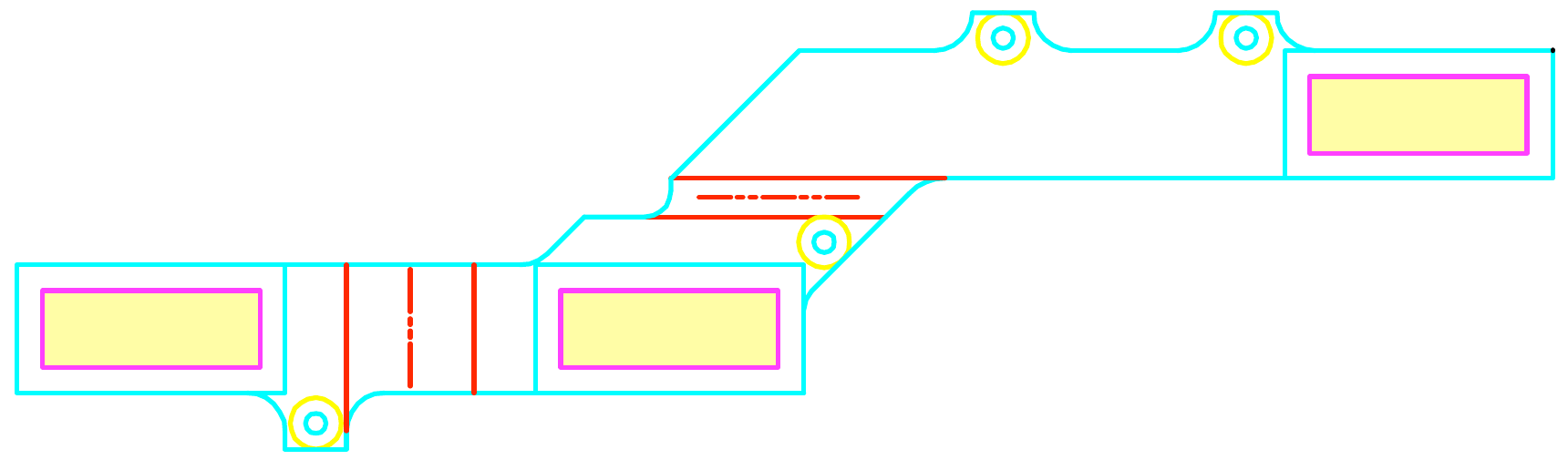} \\
   & (c)   \vspace*{-2.8cm}\\
   \includegraphics[height=5cm]{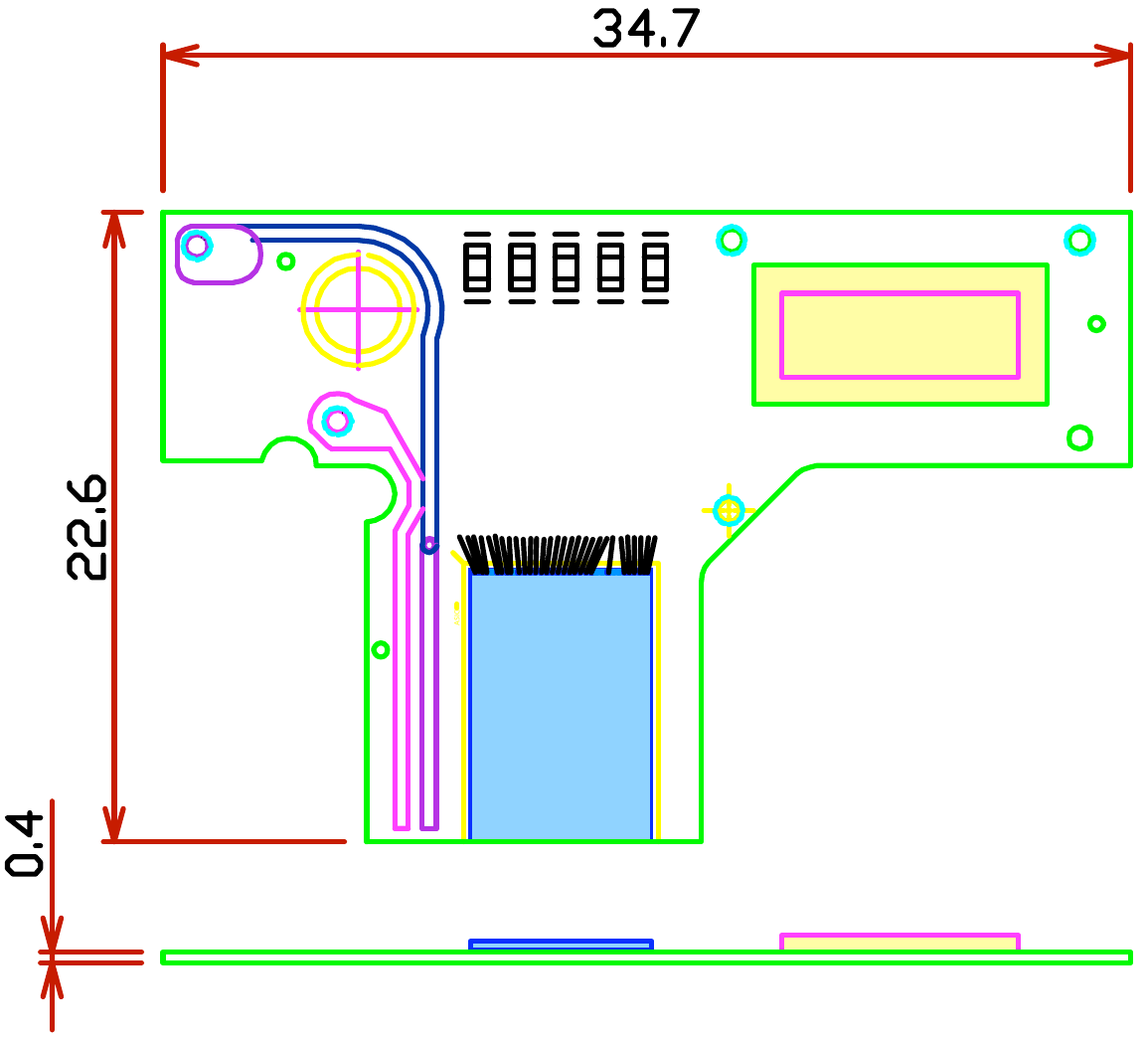} & 
   \includegraphics[height=2.0cm]{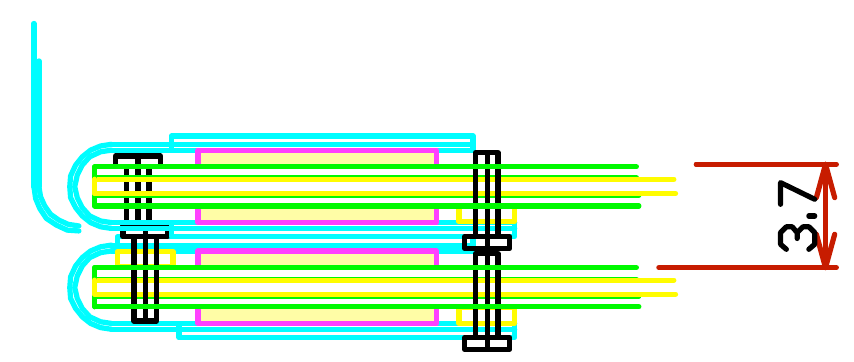} 
   \end{tabular}
   \caption{Drawings of (a) FEC (front-end card) (b) FPC (flexible printed circuits) that connects two FECs and take signals to ADB (ASIC driver board), and (c) cross-sectional view of two assemblies of two FEC and a FPC.}
   \label{fig:CC-FEC}
\end{figure}

\begin{figure}[htbp] 
   \centering
   \includegraphics[width=10cm]{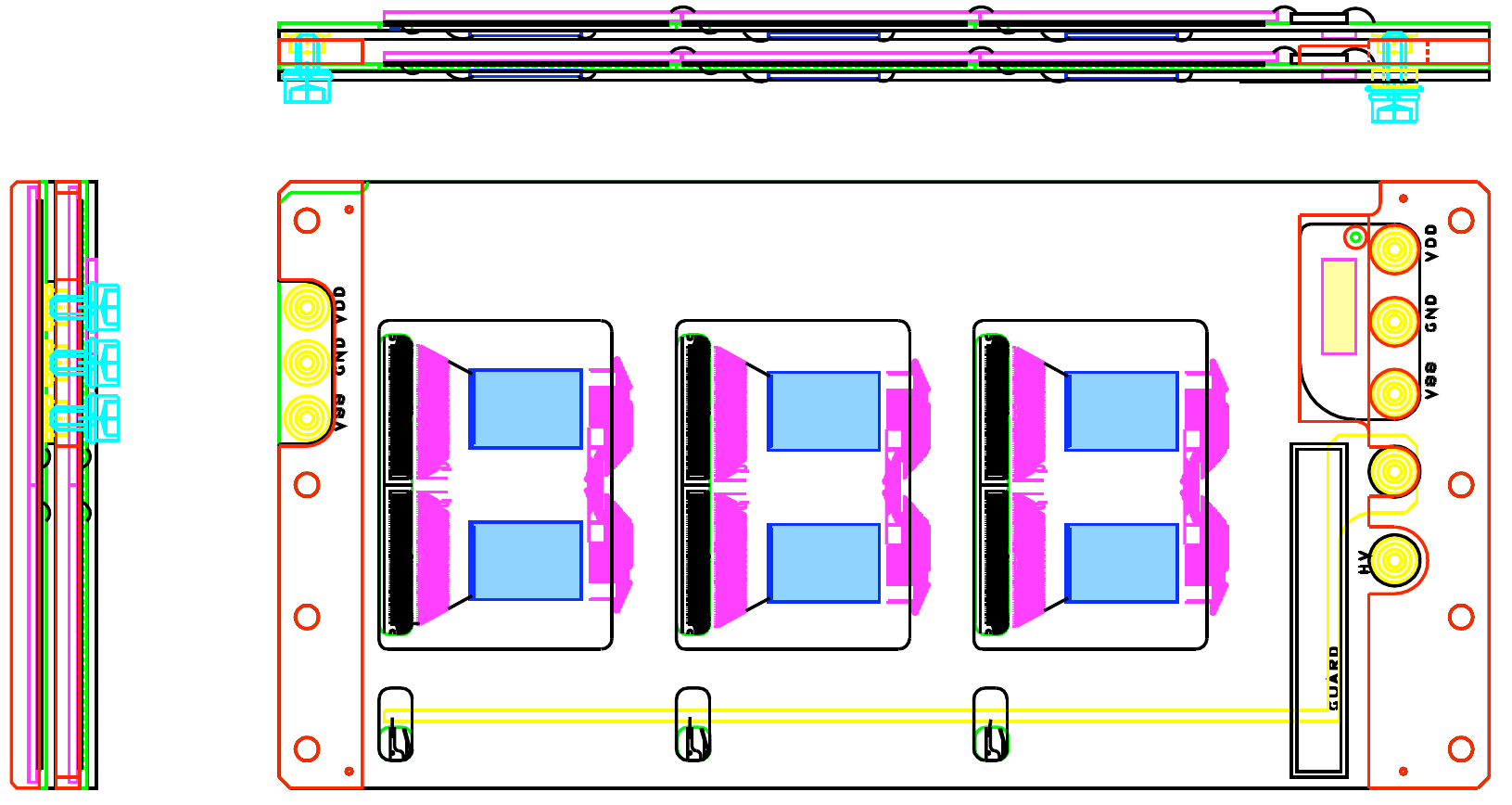} 
   \caption{Top and side views of CdTe sensor module on the side of Compton camera.}
   \label{fig:CC-side-CdTe}
\end{figure}

Each Si and CdTe sensor tray consists of Si or CdTe sensors and 
FEC (Front-End Card) mounted on both sides of the tray frame as 
shown in Figure~\ref{fig:CC-trays}.   An ASIC is mounted on each FEC.
Figure~\ref{fig:CC-FEC} shows the schematic drawings of a FEC, and 
a FPC that connects two FECs and an ADB.
Two FECs are mounted at each corner of a tray and connected by an 
FPC as shown in Figure~\ref{fig:CC-FEC} (c).

Figure~\ref{fig:CC-side-CdTe} shows drawings of side CdTe sensor module viewed from three directions.
Two CdTe sensor boards are stacked together with PEEK spacers.
A titanium frame will be attached on the back of this module to reinforce mechanical rigidity.

We have fabricated mechanical models of the Compton camera with slightly different materials (polycarbonate trays and aluminum pillars) and confirmed that those survive vibrations expected from the launch vehicle (HII-A).
We plan to fabricate a mechanical model and a thermal model of the Compton camera with the final design and evaluate mechanical and thermal properties.

\subsection{Si and CdTe sensors}
\begin{figure}[bthp] 
   \centering
   \begin{tabular}{ll}
   (a) & (b)\\
   \includegraphics[height=5cm]{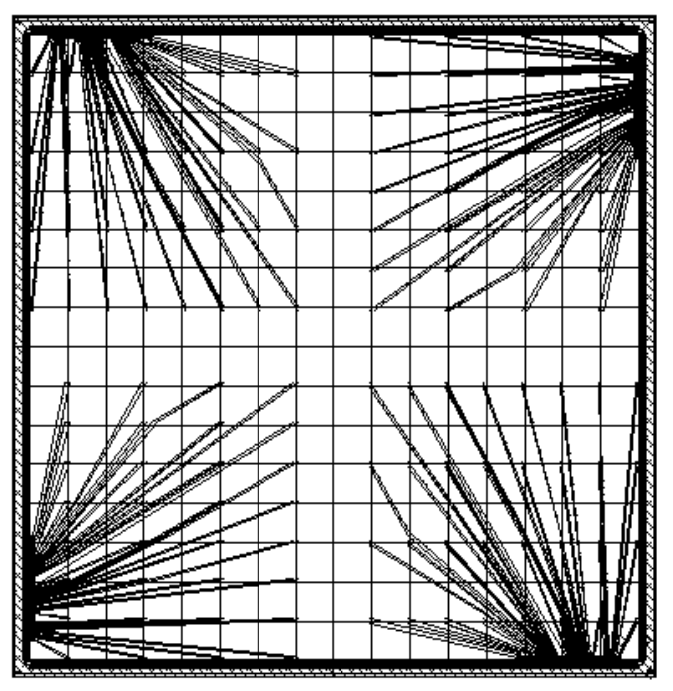} &
   \includegraphics[height=2.5cm]{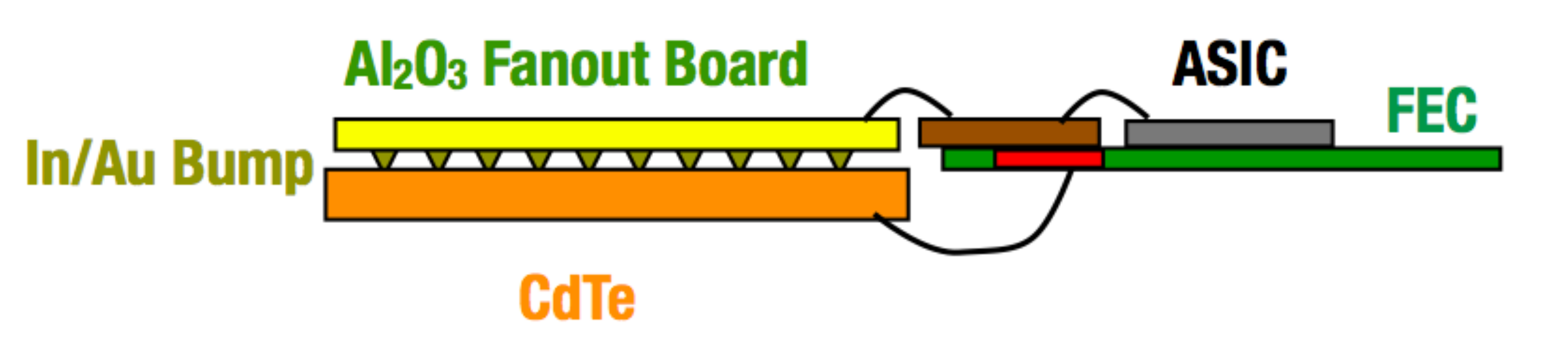} 
   \end{tabular}   
   \caption{(a) Schematic drawing of Si sensor showing layout of pixels and readout traces. (b) Conceptual illustration for the structure of a CdTe sensor module.}
   \label{fig:sensors}
\end{figure}

Si and CdTe sensors are pixellated to give two-dimensional coordinates 
with a pixel size of $3.2\times3.2$~\mmsq\ and a thickness of 0.6~mm for Si and 0.75~mm for CdTe.
Pixel size is determined to minimize the number of pixels for lower 
power consumption while avoiding the pixel size to be the dominant 
contribution to the angular resolution of Compton kinematics.
The thicknesses of Si and CdTe sensors are determined from 
constraints on the bias voltages required to operate the sensors 
at the best condition.  In order to suppress the leakage current 
from the edge of the sensor, a guard ring is placed at the periphery 
of the sensor surrounding all the pixels.  Each Si sensor has 
$16\times16$ pixels providing $5.12\times5.12$~\cmsq\ active area.
Signal of each pixel on the Si sensor is brought out to one of bonding 
pads at the corner of the sensor by a readout electrode laid out 
on top of the SiO$_2$ insulation layer with a thickness of 1.5~\micron\ 
as shown in Figure~\ref{fig:sensors} (a).  For the readout purposes, 
the Si sensors are grouped into four quadrants of $8\times8$ pixels.

A CdTe sensor has $8\times8$ pixels providing $2.56\times2.56$~\cmsq\ active area since it is difficult to fabricate the CdTe sensor much 
larger than $3\times3$~\cmsq.  CdTe sensors are tiled in an $2\times2$ 
array for each layer in the bottom and in an $2\times3$ array for each 
layer on the side to obtain the required active area.
In order to overcome small mobility and short lifetime of carriers 
in CdTe sensors, we employ a Schottky-barrier diode type CdTe sensor 
with Indium (In) anode and Platinum (Pt) cathode so that we can apply high bias voltage with low leakage current.
Indium electrode functions as a common biasing electrode while Pt 
electrodes form pixels. 
Titanium is placed on the In electrode to reduce the resistance. Gold (Au) is placed on the Pt electrode to improve connection of In/Au stud bump bonding.
Diode type CdTe sensors suffer degradation of energy resolution due to charge trapping 
over time which is called polarization. 
It is known that the polarization slows down at lower temperature and the effect of 
polarization can be reduced by applying higher bias voltage.
For example, it was found that operation of this type of CdTe 
sensor for a week has little polarization effect at $<$5\degC\ 
and $>$1000~V/mm and this polarization effect can be recovered 
by turning off the bias voltage.
Since the recovery process accelerates at a higher temperature, annealing  the sensor to 
minimize the down time may be required.

CdTe sensors cannot have integrated readout electrodes above pixel electrodes on the device unlike Si sensors.
In addition, it is not possible to wire-bond on the electrodes 
of the CdTe sensor.  In order to address those issues, we employ 
a separate fanout board to route signal from each pixel to the 
corner of the sensor where ASICs are placed.  The fanout 
board is made of 0.3~mm thick ceramic (Al$_2$O$_3$) substrate 
that allows fine pitch between electrodes to match the input 
pitch of the ASIC (91~\micron).  The CdTe sensor and the fanout board are bump-bonded via In/Au 
stud bump as shown in Figure~\ref{fig:sensors} (b).
ASIC and the fanout board are connected by wire bonding.

Table~\ref{table:sensors} summarizes specifications of Si and CdTe sensors.
\begin{table}[htdp]
\caption{Specifications for Si and CdTe sensors}
\begin{center}
\begin{tabular}{|lrr|}
	\hline 
	Description & Si & CdTe \\\hline 
	Sensor active area	 & $5.12 \times 5.12 $~\cmsq &  $2.56 \times 2.56 $~\cmsq \\ 
	Pixel area & $3.2\times 3.2$~\mmsq &  $3.2\times 3.2$~\mmsq \\ 
	Number of pixels & $16\times 16$ & $8\times 8$ \\
	Thickness of sensor & $0.62$~mm & $0.75$~mm \\ 
	Thickness of depletion (active) layer & $0.60$~mm & $0.75$~mm \\ 
	Thickness of inactive layer & $0.02$~mm & N/A \\ 
	Bias voltage & 250~V  & 1000~V\\ 
	Leakage current per pixel @ $-10$\degC & $<$50~pA & $<$50~pA \\ 
	Leakage current per pixel @ $20$\degC & $<$4000~pA & $<$4000~pA \\ 
	Width of readout electrode & 8~\micron & N/A \\ 
	Thickness of insulation for readout electrodes & 1.5~\micron & N/A\\\hline 
\end{tabular}
\end{center}
\label{table:sensors}
\end{table}%

\subsection{Application Specific Integrated Circuit}
The main performance requirements for an ASIC are low noise 
($\lesssim 2$~keV FWHM), low power ($\lesssim 0.3$~mW/channel) 
and fast readout time ($\lesssim 200\;\mu$s) to satisfy the $<$2\% 
dead time for 100~Hz trigger rate.  In order to satisfy those 
main requirements, an ASIC is developed based on the VIKING 
architecture\cite{VA94,Tajima04} which has been known for good noise performance 
and used in various space experiments like Swift, PAMELA and AGILE.

\begin{figure}[htbp] 
   \centering
   \includegraphics[height=10cm]{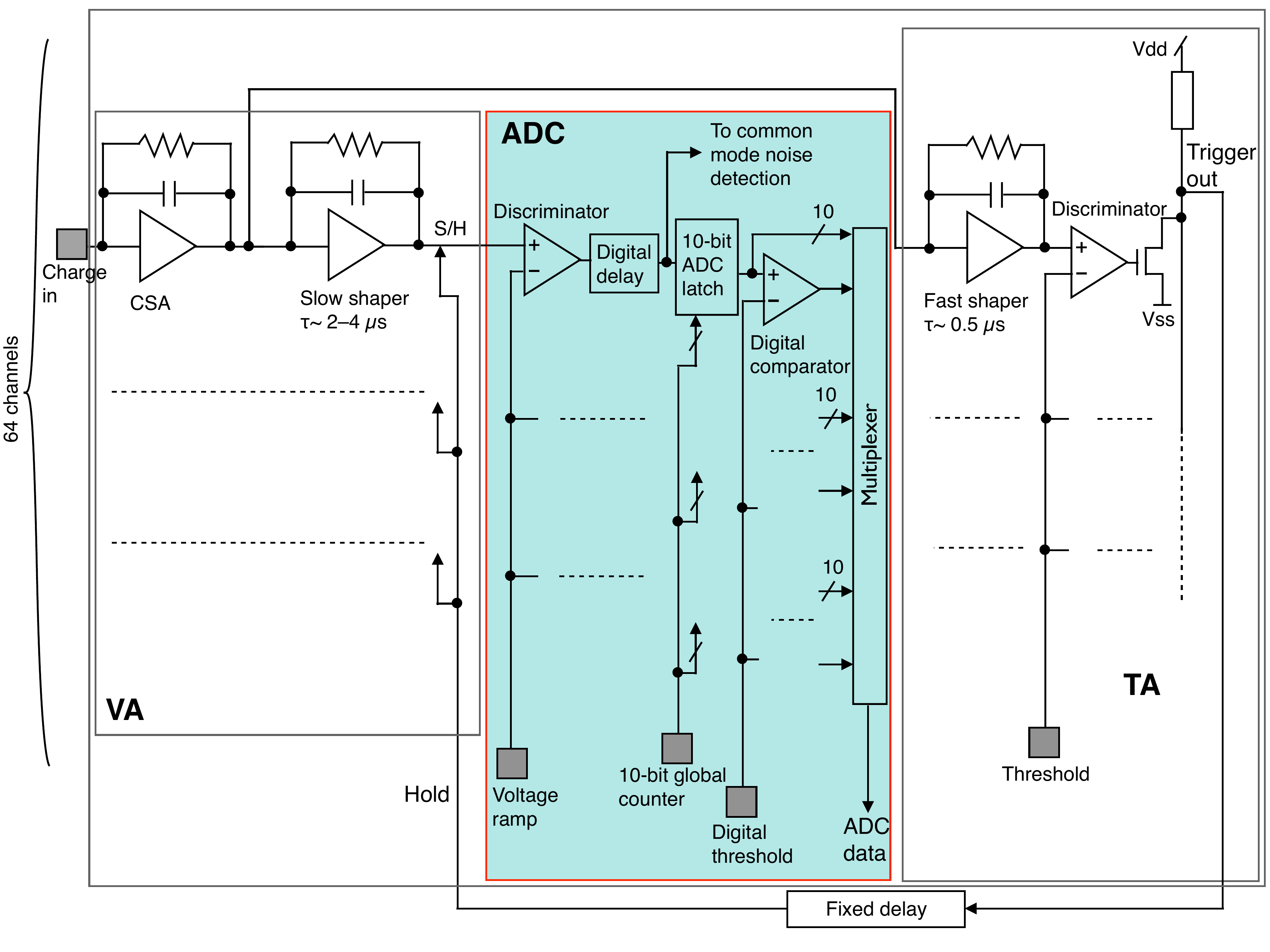} 
   \caption{Circuit diagram of the ASIC developed for the SGD. The circuits shown in a blue background are implemented in this development.}
   \label{fig:VIKING}
\end{figure}

Figure~\ref{fig:VIKING} shows the circuit diagram for the ASIC developed for the SGD (and HXI).
In the VIKING architecture ASIC, each channel consists of charge sensitive 
amplifier followed by two shapers. 
One shaper with a short shaping time is followed by a discriminator 
to form a trigger signal.
The other shaper with a long shaping time is followed by a sample and 
hold circuit to hold the pulse height at the timing specified by an external hold signal.
The hold signal is produced from the trigger signal 
with an appropriate delay.

Many important functionalities are integrated in the ASIC for the SGD in order to minimize additional components required to readout the signal as shown in the circuit diagram with a blue background region.
As a result, we only need an FPGA, several digital drivers and receivers, and passive components (resistors and capacitors) to operate 208 ASICs in a Compton camera.
The signals in all channels on the ASIC are converted to 
digital values in parallel with Wilkinson-type analog-to-digital 
converters (ADCs) where the time duration of voltage ramp to cross the 
sampled voltage is counted by a counter.
The conversion time is less than 100~$\mu$s using the external 
clock or less than 50~$\mu$s using the internal clock. (The conversion 
time depends on the pulse height of the signal.)
In order to minimize the readout time, the only channels that are read-out 
are those above a data threshold that can be digitally set for each 
channel independently from the trigger threshold.
We usually observe common mode noise from this type of ASIC at the level 
of $\sim$1~keV (can be worse if power supplies and grounding are not appropriate).
Common mode noise has to be subtracted to accurately 
apply the threshold for the zero suppression.
Common mode noise level of each event is detected by taking an 
ADC value of the 32nd (a half of number of channel) pulse height, 
corresponding to a median value of all ADC values.
With zero suppression, the readout time is $0.5\;\mu$s per ASIC when 
no data is readout and $(9+n)\;\mu$s when we readout $n$ channels. 
Without zero suppression, the readout time becomes $73\;\mu$s per ASIC. 

The ASIC produces all necessary analog bias currents and voltages 
on the chip by internal DACs (Digital to Analog Converters) except 
for the main bias current which sets the scale of all bias currents:  
this is provided by an external circuit on the FEC.
Each bit of the registers for all internal DACs and other functions 
consists of three flip-flops and a majority selector for tolerance 
against single event upset (SEU).
If the majority selector detects any discrepancies among three 
flip-flops, it will set a SEU flag which will be readout as a 
part of output data.
The ASIC is fabricated on a wafer with an epitaxial layer which 
will improve immunity against latch up.
Table~\ref{table:ASIC-spec} summarizes specifications. 
\begin{table}[htdp]
\caption{SGD ASIC (VATA450) specifications}
\begin{center}
\begin{tabular}{|l|r|}
\hline
\multicolumn{2}{|c|}{Geometrical specifications} \\\hline
Number of channels & 64 \\
Input pitch & 91~$\mu$m \\
Thickness & 0.45~mm \\ \hline
\multicolumn{2}{|c|}{Analog specifications} \\\hline
Power consumption & 0.2~mW/channel \\
Fast shaper peaking time & 0.6~$\mu$s \\
Slow shaper peaking time & $\sim$3 $\mu$s \\
Noise performance & 180~$e^-$ (RMS) at 6~pF load\\
& 1.5~keV (FWHM) for Si \\
Threshold & 1500~$e^-$ at 6~pF load \\
& 5.4 keV for Si \\
Threshold range & 625 -- 6250~$e^-$ \\
Threshold step & 208~$e^-$ \\
Dynamic range & $\pm$100,000~$e^-$ \\
& 360~keV for Si \\ \hline
\multicolumn{2}{|c|}{Digital specifications} \\\hline
ADC setup time & 5~$\mu$s \\
ADC power consumption & 0.5--2 mW/channel \\
& 5--20~$\mu$W/channel at 100~Hz \\
Data clock speed & $<$10~MHz \\
Conversion clock speed & $<$10~MHz (external clock) \\
& $<$20~MHz (internal clock)\\
Conversion time & $<$100~$\mu$s (external clock) \\
& $<$50~$\mu$s (internal clock)\\
Readout time (no data) & 0.5~$\mu$s per ASIC \\
Readout time ($n$ channels) & $(9+n)$~$\mu$s per ASIC \\ \hline
\end{tabular}
\end{center}
\label{table:ASIC-spec}
\end{table}%

The data input and output circuits on the ASIC are designed to allow 
daisy-chaining of multiple ASICs.  In one scheme, the data output 
of one ASIC can be connected to the input of another ASIC and the 
ASIC will pass the input data to the output via a shift register.
This scheme is used to set register values.
In another scheme, output of several ASICs can be connected to a single bus.
The output is controlled by passing a token from ASIC to ASIC.
Or, in the case of trigger signal, ASICs can issue trigger signals at any time since the output circuit is open-drain FET to allow multiple triggers on the same bus.
In the Compton camera, 6 or 8 ASICs are daisy chained.

\subsection{BGO active shield}
The thick active shield made of BGO scintillator is employed to reduce in-orbit background of the SGD.
The BGO crystal is heavy, and has a high stopping power, high transparency and ability to form larger crystal, while its light output is lower than NaI or CsI.
Scintillating light from a BGO is detected via an APD.

In addition to providing veto signals for cosmic rays and gamma rays from outside of the FOV, the BGO shield is also used to reduce the number of SAA (South Atlantic Anomaly) protons since those protons are the main cause of the activation of sensor materials.
The BGO shape is designed so that any trajectory that intersects with the Compton camera must go through at least 3~cm of BGO before it reaches the camera.
In order to effectively detect cosmic rays and gamma rays that interact with the BGO, the detection threshold of the BGO readout system must be lower than 100~keV.
This requirement imposes constraints on the BGO shape, reflector design, and the performance of the APD readout system.
The BGO shield consists of 30 BGO crystals and their locations and shapes are indicated by green polygons in Figure~\ref{fig:SGD-design} (a).
The weight of each BGO module is 2--6~kg and total weight is $\sim$100~kg.

\begin{figure}[htbp] 
   \centering
   \includegraphics[height=6cm]{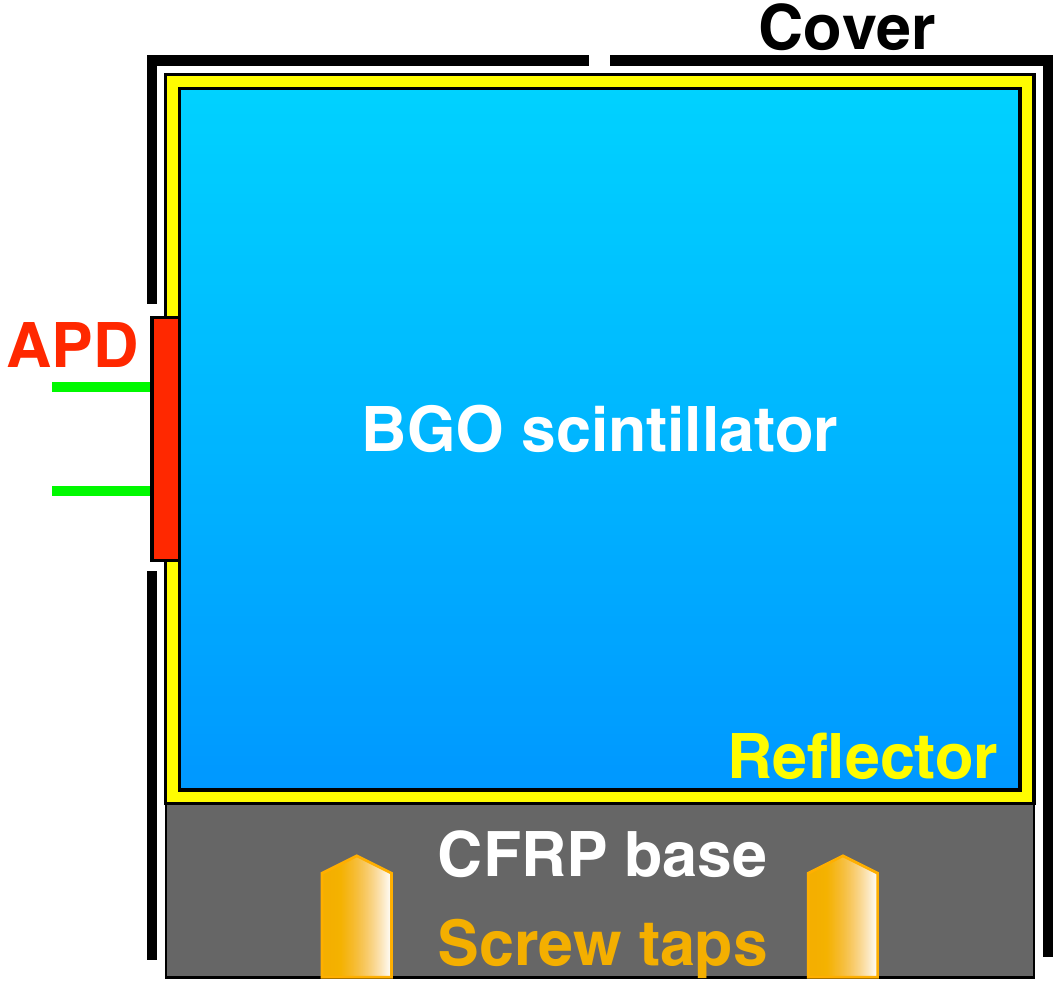} \hspace*{2cm}
   \includegraphics[height=6cm]{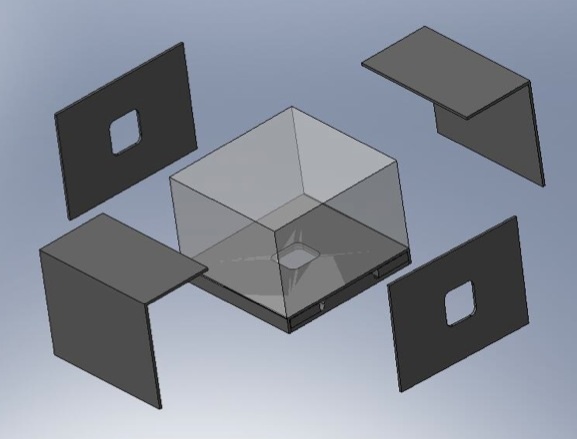}
   \caption{Conceptual views of a BGO enclosure.}
   \label{fig:BGO-concept}
\end{figure}

We employ a modular mechanical structure for the BGO shield where each BGO crystal is supported by a CFRP enclosure in order to make it easier to handle BGO modules.
The BGO enclosure consists of a CFRP base that is glued to the BGO crystal via BaSO$_4$-based reflector painted on the BGO, and CFRP covers as shown in Figure~\ref{fig:BGO-concept}.
The CFRP base screw holes to be used to attach them to the housing structure.
BaSO$_4$-based reflector is chosen for the mechanical strength that is required for the base bonding.
The remaining sides of the BGO crystal is covered by both ESR (Enhanced Specular Reflector) and Gore-Tex sheet for better reflection properties.

\subsection{Fine collimator}\label{sec:FC}
\begin{figure}[htbp] 
   \centering
   \begin{tabular}{ll}
   (a) & (b)\\
   \includegraphics[height=5.5cm]{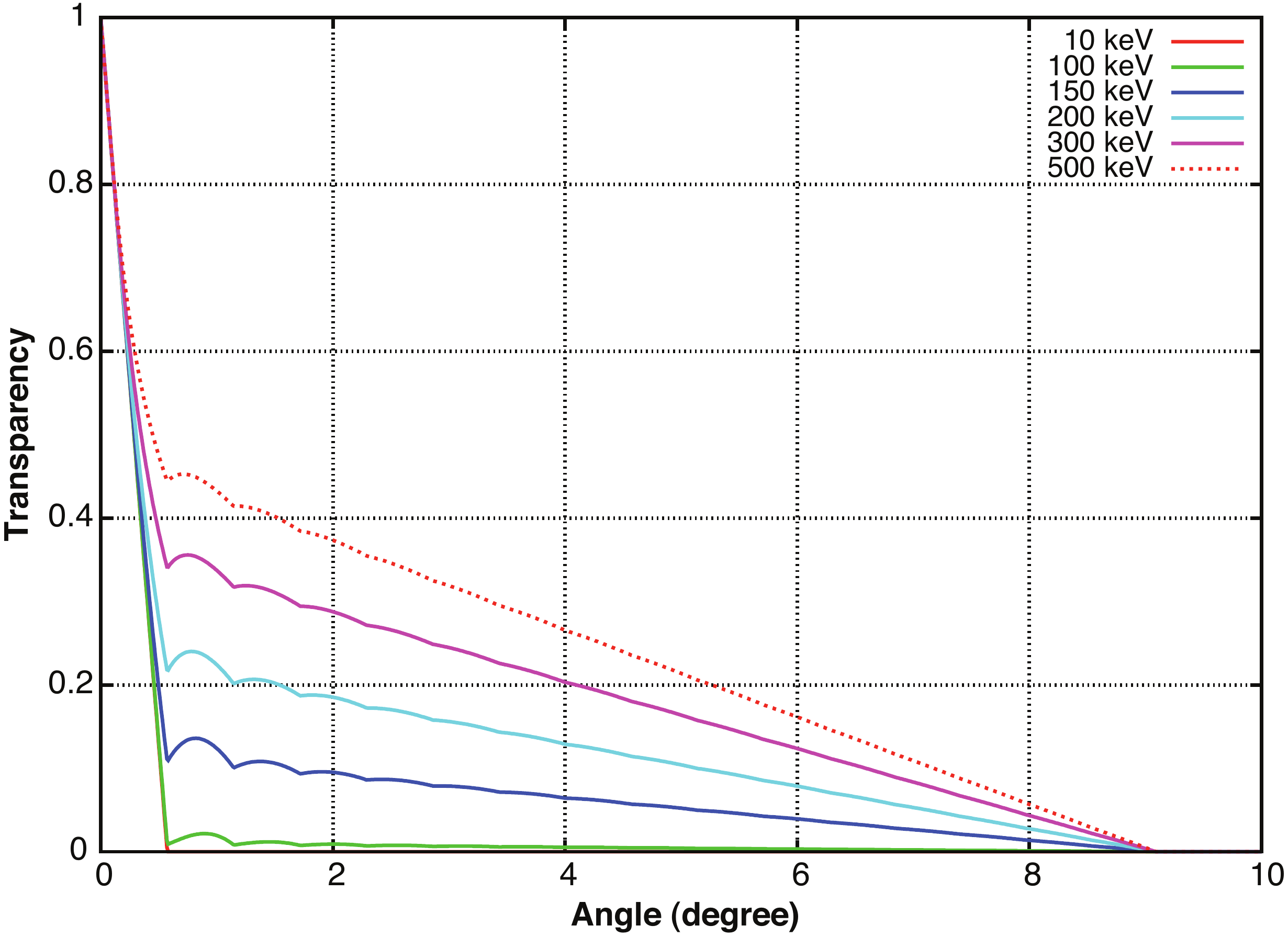} \hspace*{0.5cm} &
   \includegraphics[height=5.5cm]{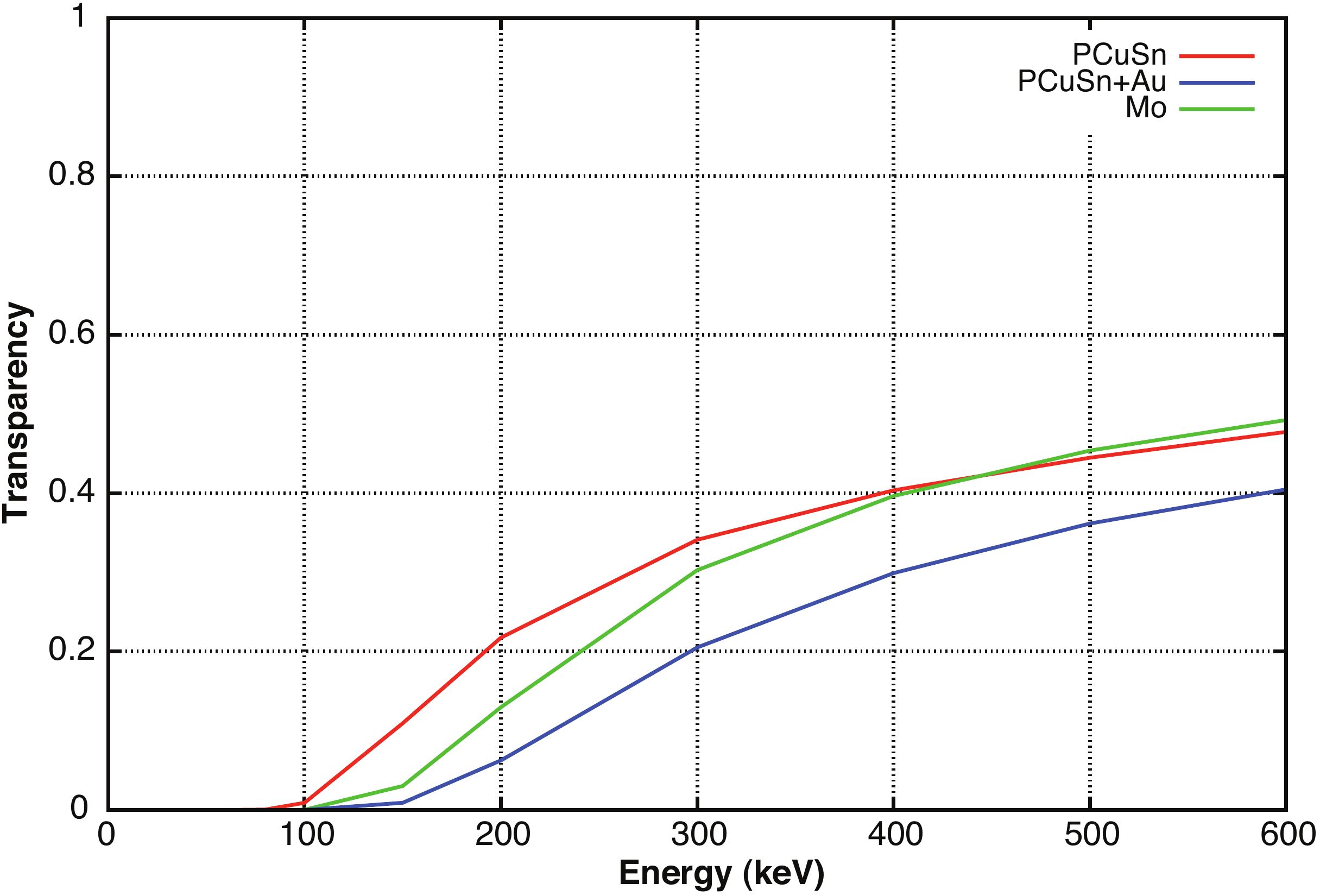}
   \end{tabular}   
\caption{(a) Analytical calculation of the transparency of the fine collimator as a function of the angle from the FOV center for the default design with 100~$\mu$m thick PCuSn.
(b) Comparison of the FC transparency among different FC materials as a function of the energy.}
   \label{fig:FC-trans}
\end{figure}

The BGO active shield has an opening of $9.7\times9.7$ deg$^2$, which is too large, resulting in CXB (cosmic X-ray backgrounds) higher than NXB (non X-ray backgrounds) and substantial source confusions within the FOV below $\sim$150~keV.
Passive collimators called fine collimators (FCs) are installed in opening of the BGO active shield, 
to reduce the FOV to 33.3~arcmin (FWHM).
Material and its thickness defines the maximum effective energy (100--150~keV) of the FC.
Note that BGOs are thick enough to detect any gamma rays in the SGD energy band ($<$600~keV).
The default choice is 0.1~mm thick PCuSn (with a length of 324~mm), which yields the maximum effective energy of $\sim$100~keV as shown in Figure~\ref{fig:FC-trans} (a).
Collimator cell size is chosen at 3.2~mm yielding an aperture opening of $\sim$94\%.
In order to ensure a transparency better than 90\%, the alignment of FC must be better than 10\% of its FOV, $\sim$3~arcmin.
Alignment mechanisms will be built into the mounting structure of the FC.

The FC material could be molybdenum (Mo) to obtain higher maximum effective energy, $\sim$150 keV as shown in Figure~\ref{fig:FC-trans} (b).
However, Mo is expected to have more activation lines than PCuSn due to higher atomic number.
We plan to have a beam test to measure the activation of Mo.

\subsection{Avalanche Photo Diode}
The APD is chosen for the photon detector of the BGO shield 
mainly due to its compact size compared with photo multipliers, 
and is chosen to be compatible with a modular structure of the shield.
Although the larger APD yields better photon collection 
efficiencies ($\propto$S$^{0.5}$), the capacitance and the 
leakage current of the APD also increases proportionally to the area.
Based on the experiments with 3, 5, 10 and 20~mm APDs, we 
concluded that 10~mm is the most appropriate for our application.
We employ HPK S8664-55 with slightly modified structure for 
less leakage current:  those were used by one of the CERN LHC 
experiments, CMS.
The APD is encapsulated by silicone resin to avoid cracks that 
appeared in epoxy resin due to thermal cycles.

Since the gain of the APD is temperature dependent ($-3$\%/\degC), 
the temperature of the APD needs to be controlled within 3\degC\ 
to keep the gain variation within 10\%.
If the temperature variation cannot be controlled within 3\degC, 
the APD bias voltage needs to be adjusted to compensate for the gain change.

Signals from the APDs are routed to CSAs (charge sensitive amplifiers) 
in shielded boxes located close vicinity of the APDs on the SGD-S housing.
Since the APD capacitance is relatively large, $\sim$250~pF, the CSA 
has to be low noise amplifier with low capacitance dependence.
The breadboard model of the CSA yields 540 electrons (FWHM) 
at 0~pF load with a capacitance dependence of 2.7 electrons/pF, 
corresponding to 1200 electrons at 250~pF.

\subsection{Electronics}
\begin{figure}[htbp] 
   \centering
   \includegraphics[width=\textwidth]{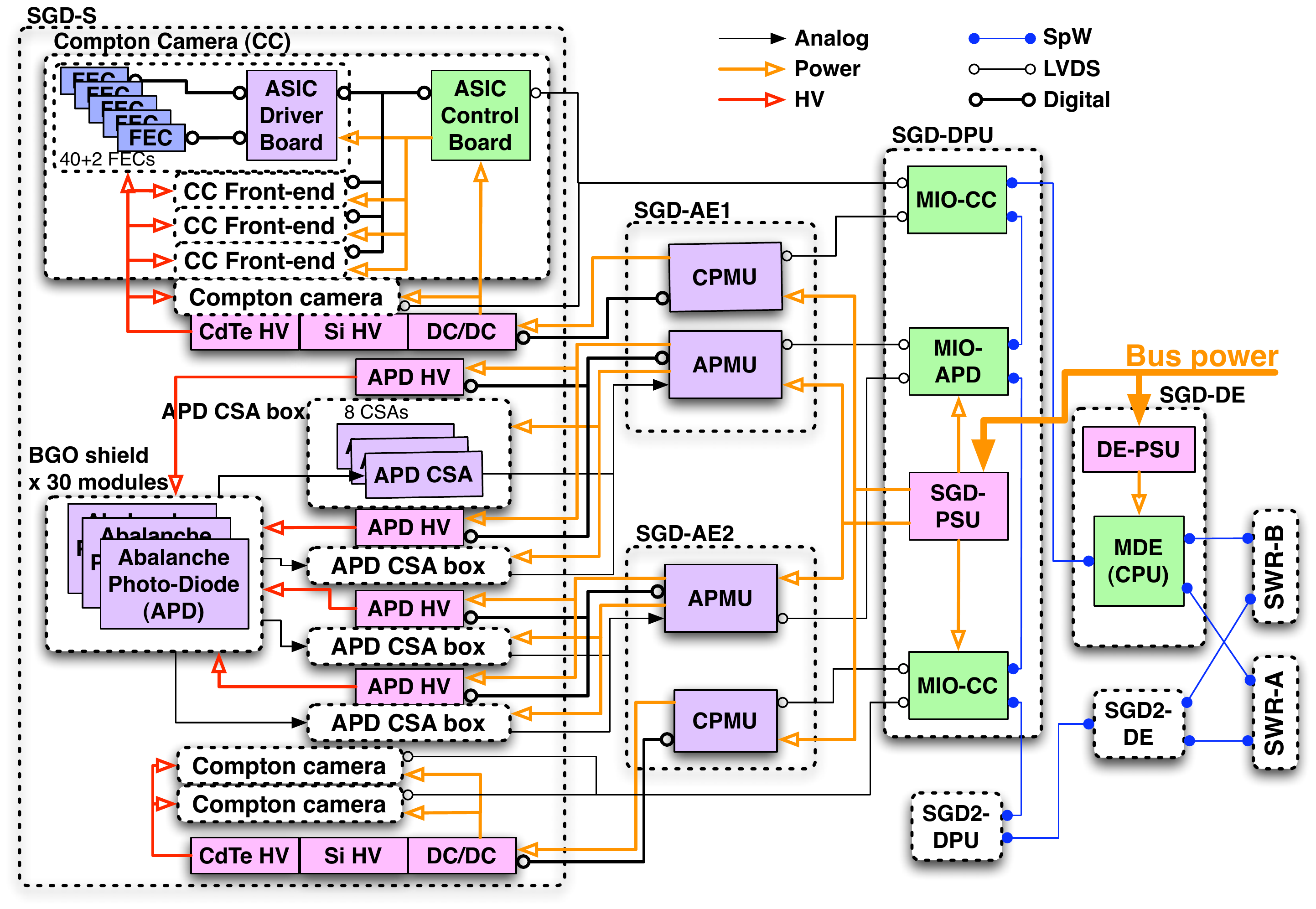} 
   \caption{Block diagram of SGD readout system.}
   \label{fig:electronics-diagram}
\end{figure}

The SGD electronics system consists of the Compton camera front-end, CPMU (Camera Power Management Unit), APD-CSA, APMU (APD Processing and Management Unit), MIO (Mission I/O) boards and power supplies as shown in a SGD electronics block diagram in Figure~\ref{fig:electronics-diagram}.

The front-end electronics of the Compton camera consists of four groups of 42 Front-End Cards (FECs) and an ASIC Driver Boards (ADBs), and an ASIC Control Board (ACB). 
Two FECs are connected back to back at the corner of each Si and CdTe trays, and are read out in daisy chain. 
FECs for the CdTe modules on the side have six ASICs that are daisy-chained on each board.
Forty FECs from the Si/CdTe trays and two FECs from the CdTe modules on the side are connected to an ADB, which is located on the side of the Compton camera.
Eight FECs (eight ASICs) are daisy-chained for Si/CdTe trays, resulting in 7 groups of ASICs for each side, 5 for Si/CdTe trays (8 ASICs each) and 2 for side-CdTe (6 ASICs each).
Only digital communication is required between ADBs and FECs and all digital signals are differential to minimize the EMI (electro magnetic interference).
Digital signals that are not used frequently are single ended between the ADB and the ACB due to constraints on the cable pin count.
The ADB detects excess current of each ASIC group in order to protect ASICs from latch-ups due to highly ironizing radiation or other origins.
We can recover ASICs from latch-ups by cycling the power supply.

ASICs are controlled by an FPGA on the ACB (one board per Compton camera). 
Functions of the ACB include: loading ASIC registers, sending a trigger to MIO and hold signals to ASICs with proper delays upon reception of triggers from ASICs, controlling analog-to-digital conversion on ASICs and data transfer from ASICs, halting data acquisition process if MIO cancels the trigger, formatting data from ASICs and send it to MIO, counting triggers and monitoring dead time.

CPMU functions include control of power switches and power supply voltages, monitor of power supply voltages and temperature.
Remote HV (high voltage) bias power supplies are controlled via slow serial data link.
We plan to ramp up and down bias power supply for the safety of the front-end electronics in the normal power up and down procedure.
However, appropriate low pass filters should be placed between each sensor and the bias power supply so that any sudden change of the HV due to unforeseen reasons does not destroy front-end electronics connected to the sensors.

The APMU receives APD signals from APD CSAs and digitize them with flash ADCs.
Digitized values are continuously monitored by an FPGA on the 
APMU.
The FPGA differentiates the APD signal and issues a trigger when 
the differentiated signal is above a certain threshold.
The time constant of the differentiation has to be optimized based on sampling frequency and the rising time of the CSA.
Slightly more sophisticated algorithm is used to calculate more accurate pulse height information given the trigger timing of the Compton camera.
Other APMU functions include HK such as control of power switches and power supply voltages, monitor of power supply voltages and temperature.
Remote HV power supplies for APD are controlled via slow serial data link.

MIO functions include: recording event time tag, assemble veto information upon reception of trigger signals from Compton camera, sending trigger cancel within 10~$\mu$s if necessary based on veto information, managing dead time, veto signals from APMU and ASIC registers which includes checking SEU bit from ASIC data, formatting data including sensor data, time tag, veto hit pattern, and send them to MIO, and controlling CPMU including reception of HK data from CPMU.

Communications between the Compton camera and MIO, and that between CPMU/APMU and MIO are handled via 3-line (CLK, DATA, STRB) serial protocol on LVDS physical layer.
We have two additional real-time LVDS lines dedicated for trigger and trigger acknowledgement signals  between CPMU and MIO.
We also have dedicated LVDS lines to issue two types of veto signals between APMU and MIO.

%
\section{Expected Scientific Performance}
\begin{figure}[bth]
   \centering 
   \begin{tabular}{ll}
   (a) & (b) \\
   \includegraphics[height=5.4cm]{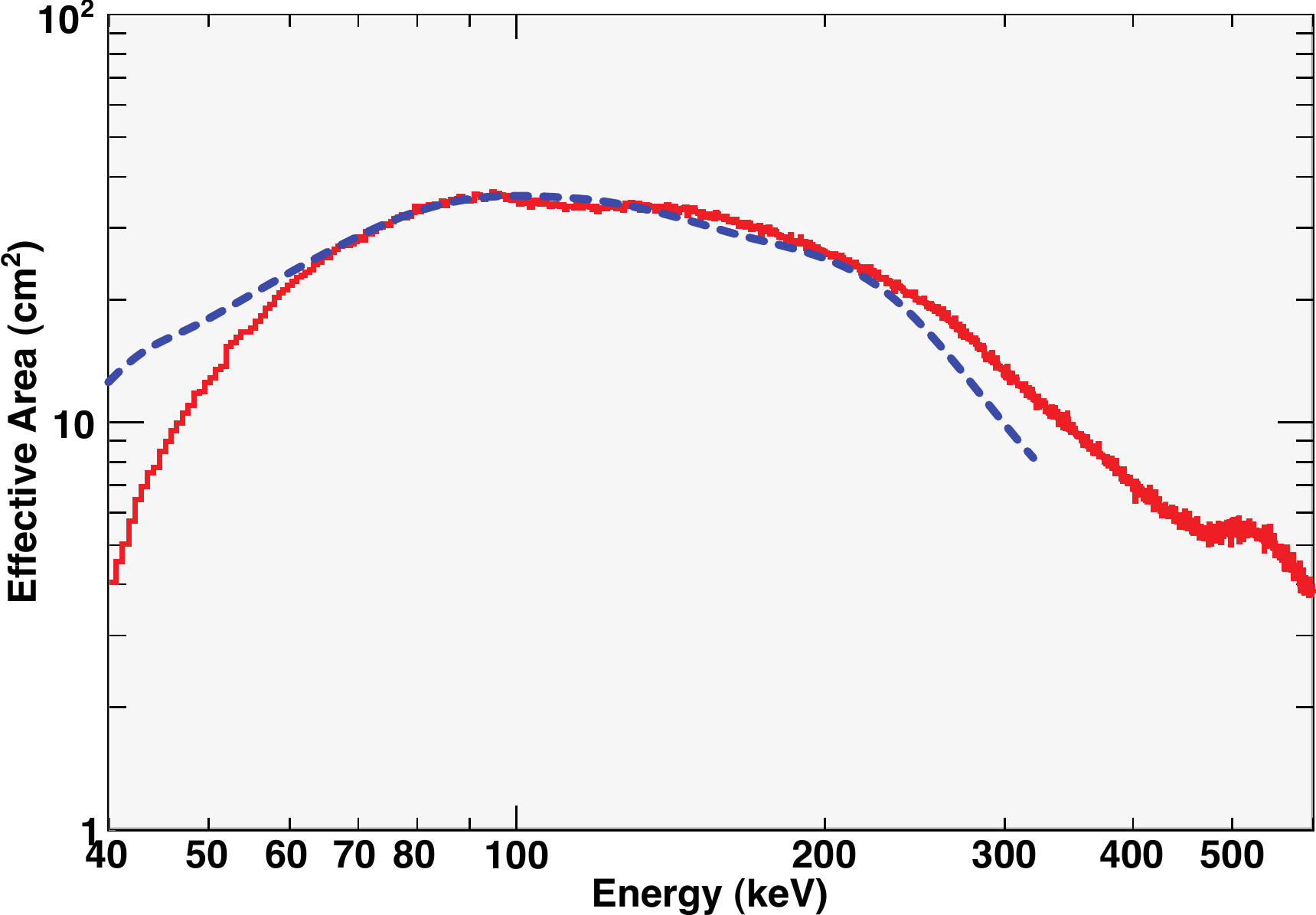} \hspace*{0.5cm} &
   \includegraphics[height=5.4cm]{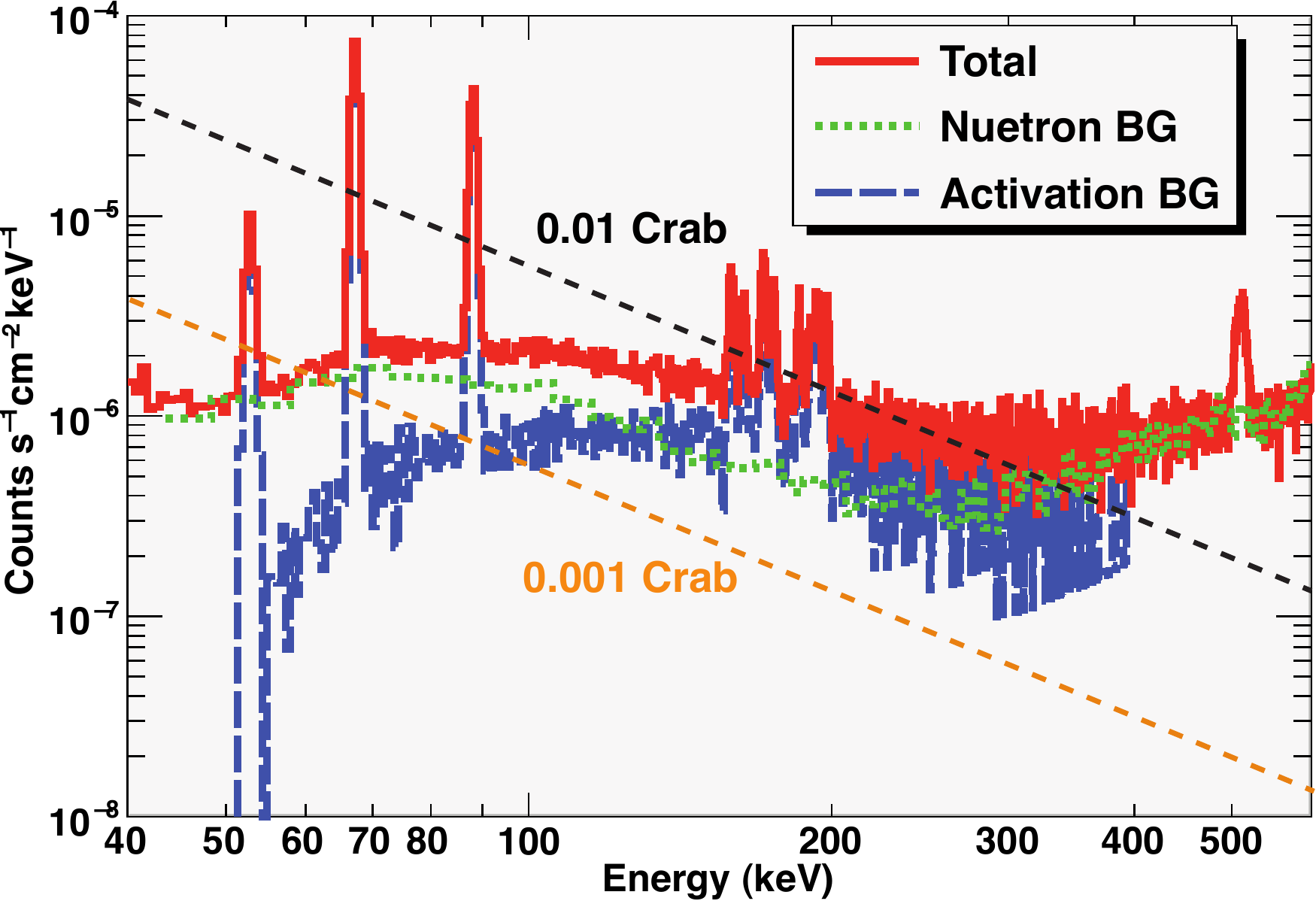} 
   \end{tabular}
   \caption{(a) Effective area (red solid) and inverse of MDP in arbitrary unit (blue dashed) as a function of incident energy. (b) Background flux as a function of reconstructed energy.}
   \label{fig:SGD-performance}
   \end{figure}

Effective area, non X-ray backgrounds, and sensitivities are evaluated by Geant4-based Monte Carlo simulations. 
The solid line in Figure~\ref{fig:SGD-performance} (a) shows the effective area as a function of the incident energy for the current SGD design.
Maximum effective area of more than 30~\cmsq\ is realized at around 80--100~keV, which corresponds to $\sim$15\% reconstruction efficiency since the geometrical area of the SGD is 210~\cmsq.
The effective area at low energies is suppressed due to the photo-absorption in Si while the loss at high energies is due to multiple-Compton events, which can be recovered by improved reconstruction algorithm.
The dotted line in Figure~\ref{fig:SGD-performance} (a) shows the inverse of minimum detectable polarization (MDP) in arbitrary units assuming no background.
The polarization sensitivity falls off slower at low energies and faster at high energies due to lower modulation factor resulting from more forward scattering at higher energies.
This result indicates that SGD is sensitive to the polarization in the 50--200~keV energy band.
   \begin{figure}[bth]
   \centering
   \begin{tabular}{ll}
   (a) & (b) \\
   \includegraphics[height=5.7cm]{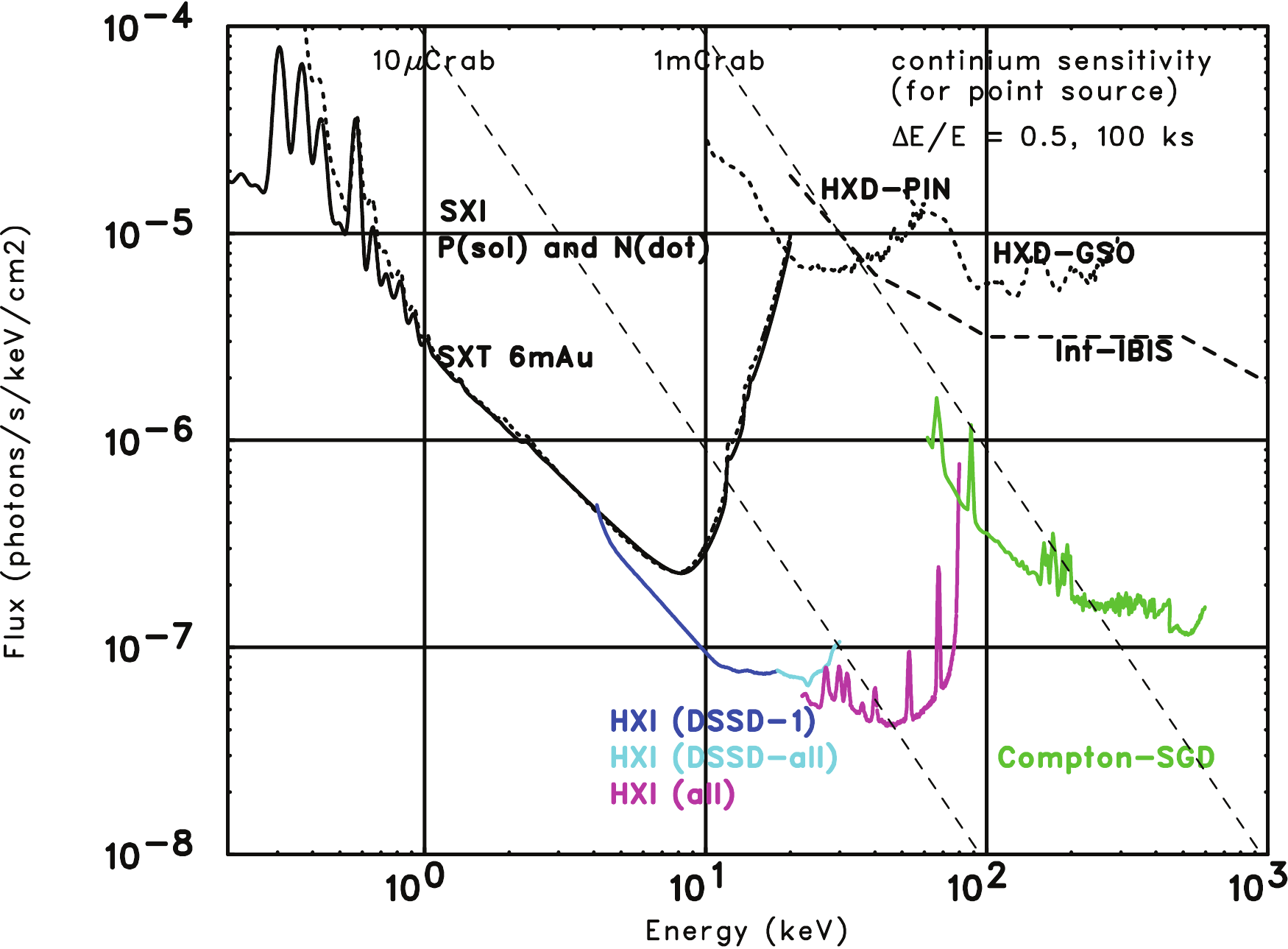} \hspace*{0.5cm} &
   \includegraphics[height=5.7cm]{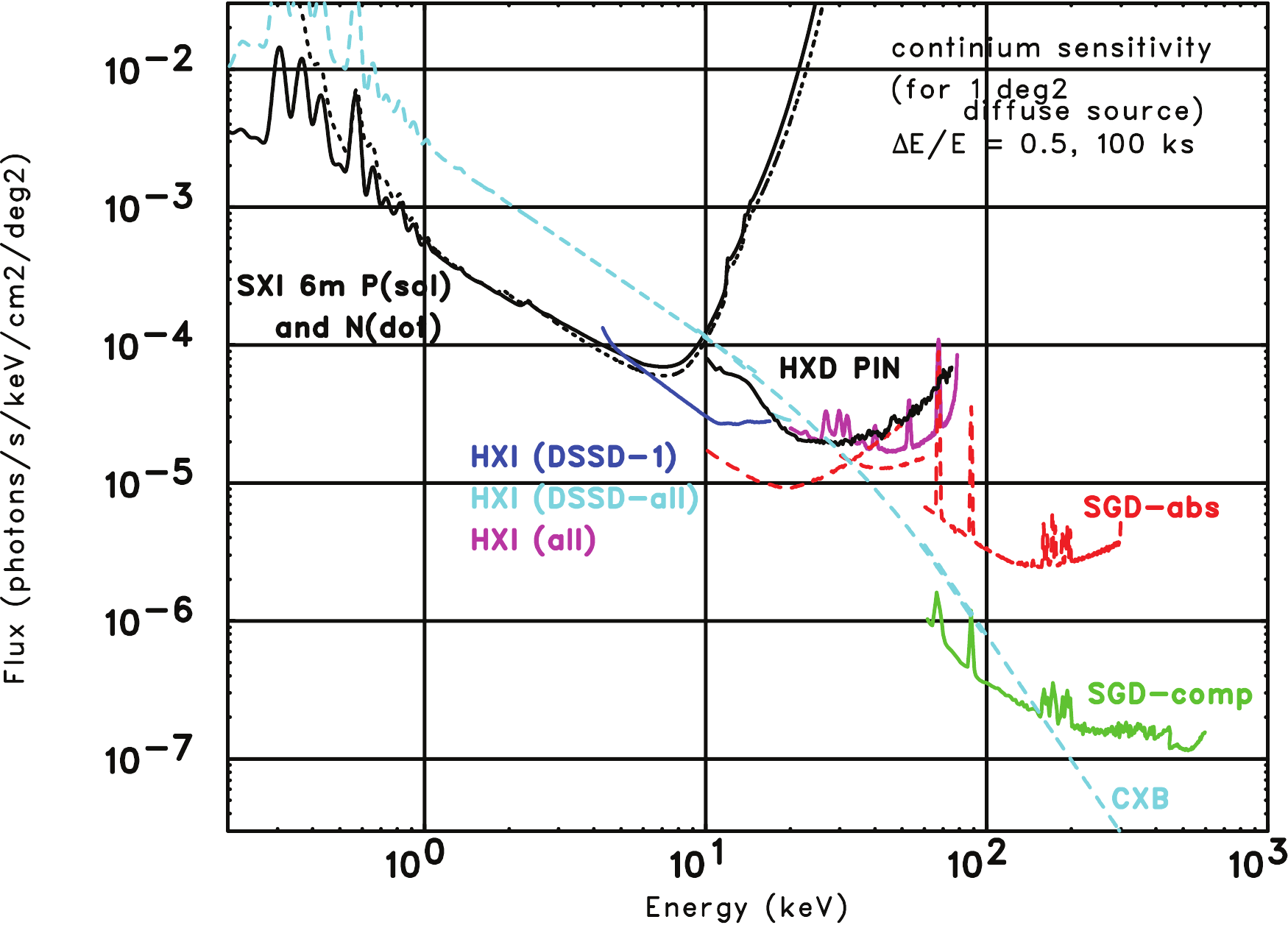} 
   \end{tabular}
   \caption{3$\sigma$ sensitivity targets for the SXI, HXI and SGD in the ASTRO-H mission for continuum emissions from (a) point sources and (b) extended sources, assuming an observation time of 100~ks and comparison with other hard X-ray and soft gamma-ray instruments.}\label{fig:sensitivity}
   \end{figure} 

Main in-orbit background components of the SGD are expected 
to be activations induced during the SAA passages and 
elastic scatterings of albedo neutrons, at the expected 
orbit of ASTRO-H (altitude of 550 km with an inclination angle of 31\degree). 
These background events can be heavily suppressed 
by a combination of multi-layer low-$Z$/high-$Z$ sensor configuration, active 
shield, and the background rejection based 
on the Compton kinematics.
The remaining background level is estimated to be much lower 
than any past instrument as shown in Figure~\ref{fig:SGD-performance} (b).
The neutron background (green dotted curve) is estimated 
by the simulation assuming the neutron spectrum described in Ref.~\citenum{Armstrong73}.
The flux of the neutron background is scaled by a 
factor of two based on the background studies of the 
Suzaku hard X-ray detector \cite{Fukazawa09}.
The spectrum of the activation background (blue dashed 
curve) is estimated from experimental results on the 
radioactivities induced by mono-energetic protons \cite{Murakami03}.
The flux is scaled by a rejection factor expected from 
constraints by the Compton kinematics.
The signal fluxes corresponding to 1/100 and 1/1000 of 
the Crab brightness are overlaid in black and orange dotted 
straight lines, respectively.
This clearly illustrates that the expected background in SGD 
varies from 1/1000 to 1/100 of the Crab brightness in the 50--400~keV band.
Fig.~\ref{fig:sensitivity} shows 3$\sigma$ sensitivity for three 
instruments in the ASTRO-H mission, the SXI, HXI and SGD, for continuum emission from (a) 
point sources and (b) extended sources ($1\degM\times 1\degM$) with 
an observation time of 100~ks and comparison with other instruments. 
(Sensitivity depends on the bandwidth of each 
point and observation time, and can be lower than the background 
level with sufficient statistics.)
SGD represents great improvement in the soft gamma-ray band 
compared with the currently operating INTEGRAL\cite{INTEGRAL} 
or Suzaku HXD, and extends the bandpass to well above the cutoff 
of hard X-ray telescopes, which in turn allows us to study the 
high energy end of the particle spectrum.
Combined with the SXI and the HXI on board the ASTRO-H, SGD realizes 
unprecedented level of sensitivities from soft X-ray to soft gamma-ray band.

\begin{figure}[bthp] 
   \centering
   \begin{tabular}{ll}
\vspace*{-0.cm}
   (a) & (b) \\
   \includegraphics[height=6cm]{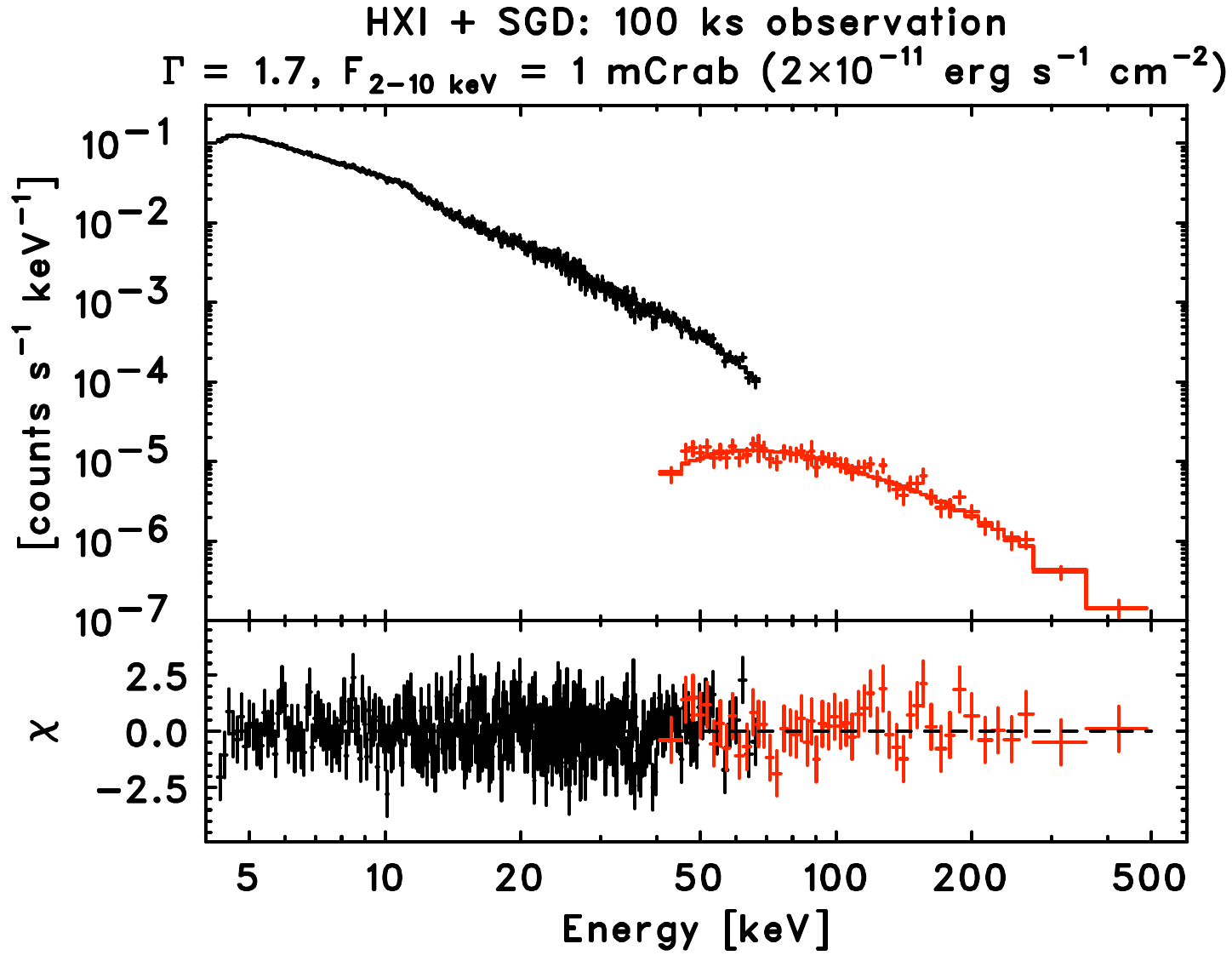} \hspace*{0.5cm}&
   \includegraphics[height=6cm]{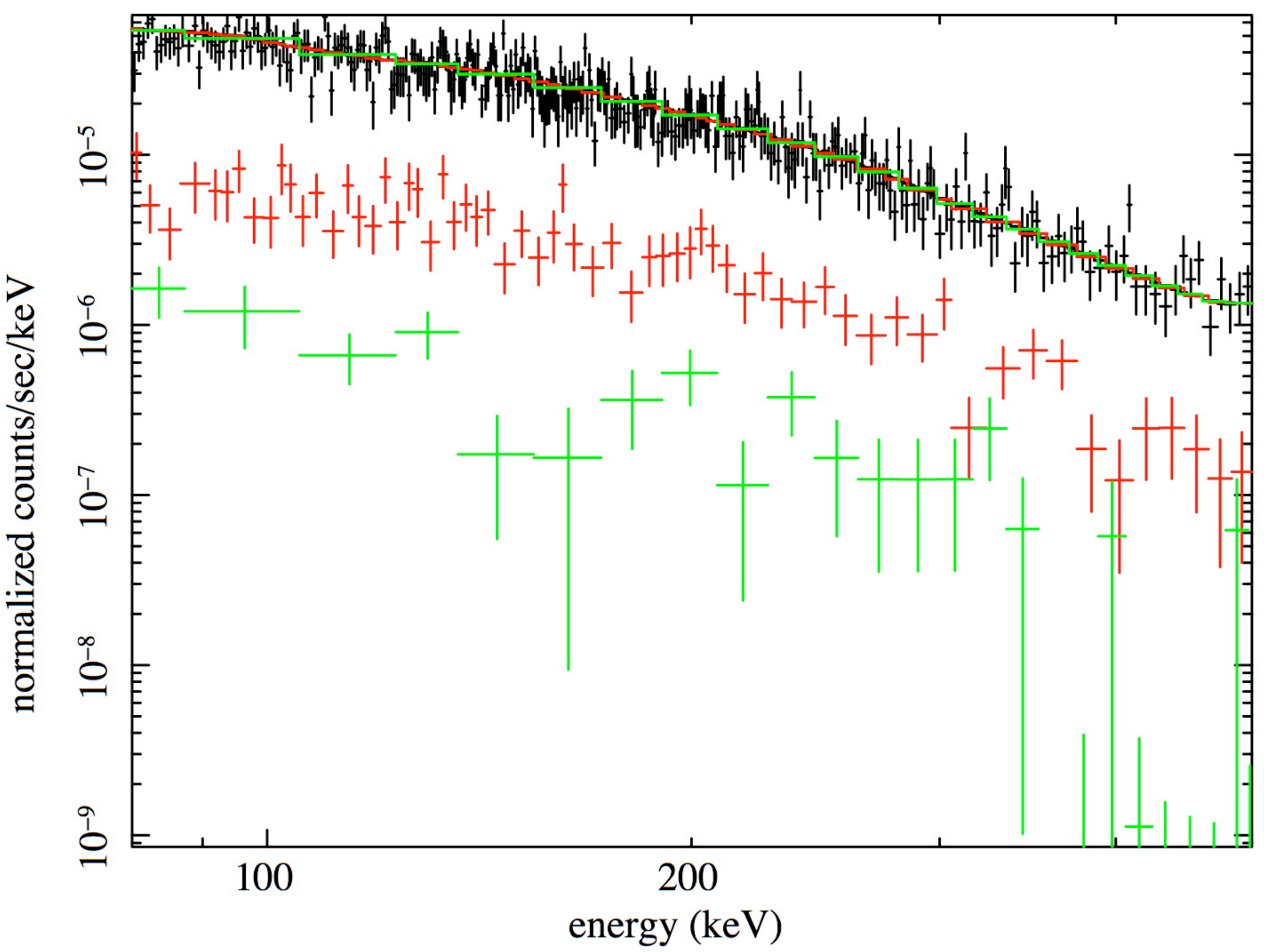}
   \end{tabular}
   \caption{(a) HXI (black) and SGD (red) simulation results for a 100~ks observation of a source with 1/1000 of the Crab brightness and power law index of 1.7. 
   (b) SGD simulation results for a 100~ks observation of bremsstrahlung emissions from Cas~A with three magnetic field hypotheses, 0.1~mG (black), 0.3~mG (red) and 1.0~mG (green).}
   \label{fig:SGD-science}
\end{figure}
   \begin{figure}[bth]
   \centering
   \begin{tabular}{ll}
   (a) & (b) \\
   \includegraphics[height=6cm]{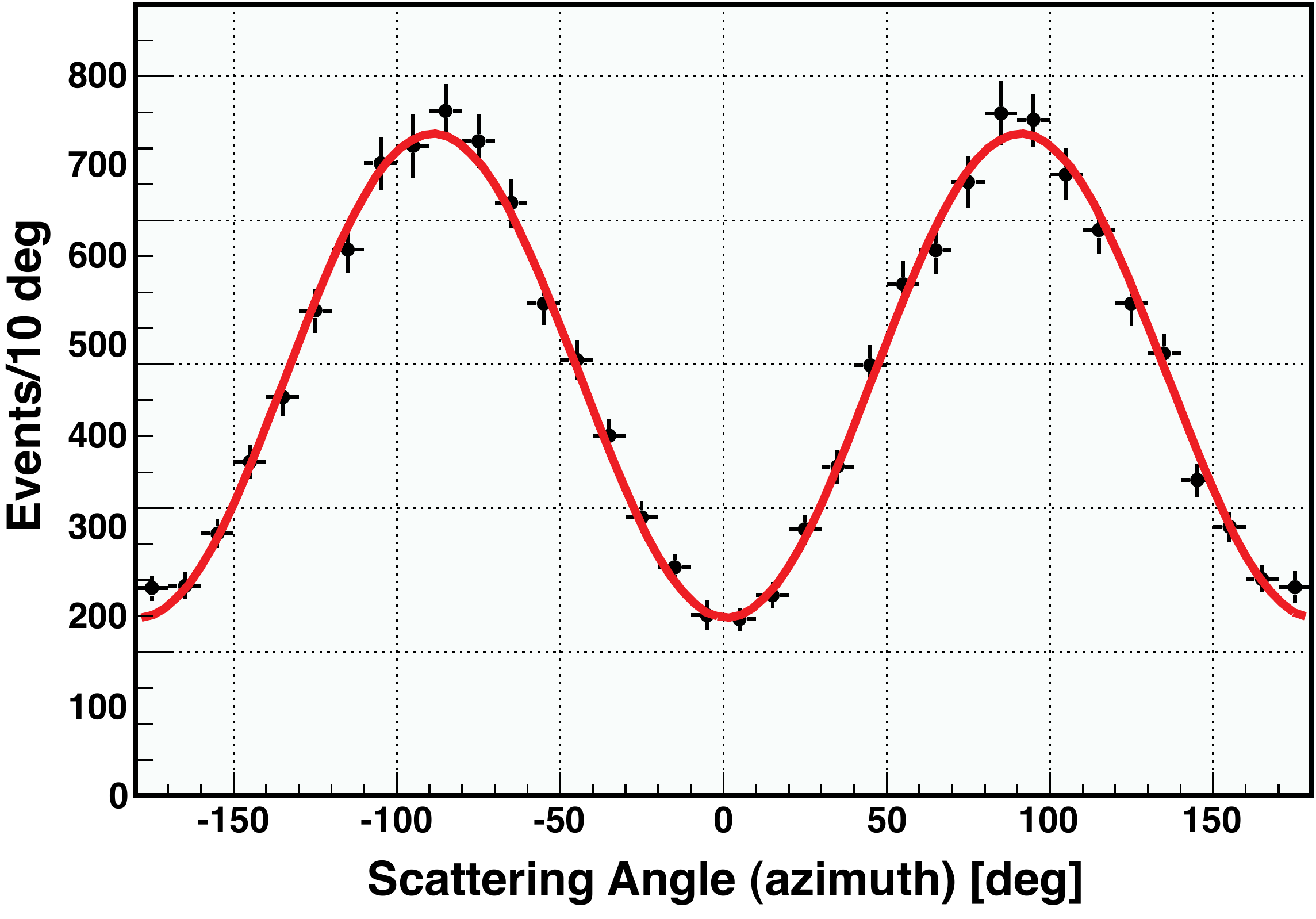} \hspace*{1.0cm} &
   \includegraphics[height=6cm]{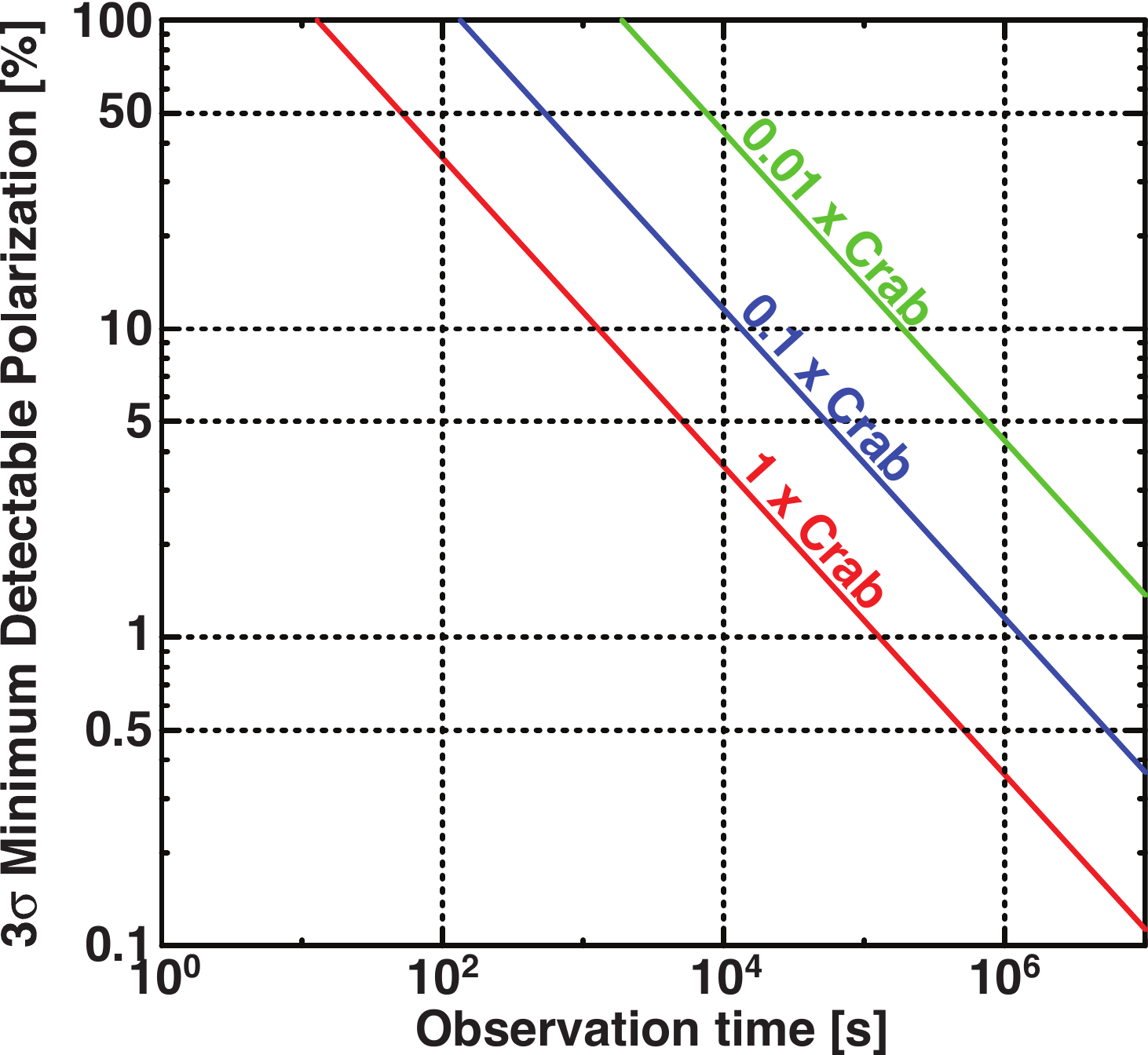}
   \end{tabular}
   \caption{(a) Efficiency-corrected azimuth angle distribution of Compton scattering from a source with a brightness of Crab and 100\% linear polarization in a 10~ks observation. (b) $3\sigma$ MDP as a function of observation time for sources with 1, 1/10 and 1/100 of the Crab brightness.} 
   \label{fig:pol-performance}
   \end{figure} 

A simulation results shown in Figure~\ref{fig:SGD-science} (a) demonstrate 
that spectral index can be measured within 10\% error for a 100~ks observation 
of a source with 1/1000 of the Crab brightness in 2--10~keV and power law index 
of 1.7 using the current SGD design parameters. 
Another type of SGD target is supernova remnants where we can 
study nature of particle accelerations.
Cas~A is one of the most promising SNR for the 
SGD since sizable non-thermal bremsstrahlung emission is expected in the SGD band.
Figure~\ref{fig:SGD-science} (b) shows simulation results 
for observation of non-thermal bremsstrahlung from a supernova 
remnant, Cas~A, which confirms that SGD can determine the 
magnetic field of Cas~A with a 100~ks observation.
This measurement will have significant implications on modeling 
of multi-wavelength observations since a model with leptonic 
origin predicts $B\approx0.12$~mG while a hadronic model prefers $B\approx0.5$~mG.

The polarization signature of incident gamma-ray is detected by 
the modulation of the azimuth angle distribution of Compton 
scattering  in SGD as shown in Figure~\ref{fig:pol-performance}~(a) 
for a 100\%-polarized source.
A fit to $AVG[1+Q\cos2(\phi-\chi_0)]$ yields $Q=56.7\pm1.0$\%, 
where $Q$ is the modulation factor which is proportional to the 
polarization degree and $\chi_0$ is the angle of the polarization vector.
Using the modulation factor obtained here and the background level 
described above, we can calculate the MDP (minimum detectable 
polarization) analytically assuming no systematic effect from 
uneven backgrounds and uncertainties of the detector response.
Figure~\ref{fig:pol-performance}~(b) shows the $3\sigma$ MDP 
as a function of the observation time for sources with 1, 1/10 
and 1/100 of the Crab brightness, which can be parametrized 
as $3.5\%\sqrt{10^4/t_\mathrm{obs}}$, $3.6\%\sqrt{10^5/t_\mathrm{obs}}$ 
and $4.3\%\sqrt{10^6/t_\mathrm{obs}}$, respectively, where 
$t_\mathrm{obs}$ is the observation time in seconds.
We can conclude that SGD can detect polarization from sources 
down to a few$\times1/100$ of the the Crab brightness with a 
polarization degree of several \% in a few$\times100$~ks of observation time.

%
\section{Summary}
The Soft Gamma-ray Detector (SGD) onboard the next Japanese X-ray astronomy 
satellite ASTRO-H is designed to measure spectra of celestial sources with $>$1/1000 
of Crab brightness in the 40--600~keV energy band, which is the highest end 
of the ASTRO-H energy coverage. The sensitivity of the SGD presents more than 
an order of magnitude improvement in the soft gamma-ray band as compared with 
the currently operating INTEGRAL or Suzaku HXD instruments. Combined with the 
soft and hard X-ray imagers (SXI and HXI) on board of ASTRO-H, the SGD realizes 
unprecedented level of sensitivities from soft X-ray to soft gamma-ray band.
A key to achieve such sensitivity is a low background realized by a combination 
of the Compton camera surrounded by the BGO active shield where the incoming 
photon angle constrained by Compton kinematics is required to be consistent 
with the narrow field view of the active shield and passive collimator.
The SGD is also capable of measuring polarization of celestial 
sources brighter than a few $\times 1/100$ of the Crab Nebula, polarized 
above the $\sim$10\%. This capability is expected to yield polarization 
measurements in several celestial objects, providing new insights into 
properties of soft gamma-ray emission processes.

The combination of low-$Z$ (Si) and high-$Z$ (CdTe) sensors allows us to employ 
appropriate sensor materials to lower the energy threshold, to minimize 
Doppler broadening while maximizing absorption efficiencies of the 
scattered photons.  The low-$Z$/high-$Z$ arrangement also suppresses contributions 
from neutron and activation backgrounds.

The SGD successfully completed preliminary design review in May 2010 and 
is currently in the detailed design phase.  The ASTRO-H is expected to be 
launched in early 2014.



\bibliography{mybib}   

\begin{thebibliography}{10}

\bibitem{NeXT08}
Takahashi, T. et~al., ``The {NeXT} {X}-ray mission, new exploration {X}-ray
  telescope,'' in [{\em UV to Gamma-Ray Space Telescope
  Systems}{\nolinebreak\hspace{0.1em}]},  {\em SPIE} {\bf 7011},  14T (2008).

\bibitem{Takahashi10}
Takahashi, T., Mitsuda, K., Kelley, R.~L., et~al., ``The {ASTRO-H} mission,''
  in [{\em Space Telescopes and Instrumentation 2010: Ultraviolet to Gamma
  Ray}{\nolinebreak\hspace{0.1em}]},  {\em SPIE} {\bf 7732} (2010).

\bibitem{Takahashi02-NeXT}
Takahashi, T., Kamae, T., and Makishima, K., ``Future hard {X}-ray and
  gamma-ray observations,'' in [{\em New Century X-ray Astronomy, ASP
  (Astronomical Society of the Pacific Conference
  Series)}{\nolinebreak\hspace{0.1em}]},   {\bf 251},  210--213 (2002).

\bibitem{Takahashi02}
Takahashi, T., Nakazawa, K., Kamae, T., Tajima, H., Fukazawa, Y., Nomachi, M.,
  and Kokubun, M., ``High resolution {CdTe} detectors for the next generation
  multi-{Compton} gamma-ray telescope,'' in [{\em X-ray and Gamma-ray
  Telescopes and Instruments for Astronomy,
  {SPIE}}{\nolinebreak\hspace{0.1em}]},  Truemper, J.~E. and Tananbaum, H.~D.,
  eds.,  {\bf 4851},  1228--1235 (2002).

\bibitem{Takahashi03-SGD}
Takahashi, T., Makishima, K., Fukazawa, F., Kokubun, M., Nakazawa, K., Nomachi,
  M., Tajima, H., Tashiro, M., and Terada, Y., ``Hard {X}-ray and {Gamma}-ray
  detectors for the {NeXT} mission,'' {\em New Astro. Rev.}~{\bf 48},  309--313
  (2004).

\bibitem{Takahashi04-SGD}
Takahashi, T., Awaki, A., Dotani, T., Fukazawa, Y., Hayashida, K., Kamae, T.,
  Kataoka, J., Kawai, N., et~al., ``Wide-band {X}-ray imager ({WXI}) and soft
  gamma-ray detector ({SGD}) for the {NeXT} mission,'' in [{\em UV and
  Gamma-Ray Space Telescope Systems}{\nolinebreak\hspace{0.1em}]},  {\em SPIE}
  {\bf 5488},  549--560 (2004).

\bibitem{HXI08}
Kokubun, M. et~al., ``Hard {X}-ray imager {HXI} for the {NeXT} mission,'' in
  [{\em UV to Gamma-Ray Space Telescope Systems}{\nolinebreak\hspace{0.1em}]},
  {\em SPIE} {\bf 7011},  21K (2008).

\bibitem{Tajima05}
Tajima, H., Kamae, T., Madejski, G., Mitani, T., Nakazawa, K., Tanaka, T.,
  Takahashi, T., Watanabe, S., et~al., ``Design and performance of the {Soft
  Gamma-ray Detector} for the {NeXT} mission,'' {\em IEEE Trans. Nucl.
  Sci.}~{\bf 53},  2749--2757 (2005).

\bibitem{Takahashi01b}
Takahashi, T. and Watanabe, S., ``Recent progress in {CdTe} and {CdZnTe}
  detectors,'' {\em IEEE Trans. Nucl. Sci.}~{\bf 48},  950--959 (2001).

\bibitem{Watanabe05}
Watanabe, S., Tanaka, T., Nakazawa, K., Mitani, T., Oonuki, K., Takahashi, T.,
  Takashima, T., Tajima, H., Fukazawa, Y., Nomachi, M., Kubo, S., Onishi, M.,
  and Kuroda, Y., ``A {Si/CdTe} semiconductor {Compton} camera,'' {\em IEEE
  Trans. Nucl. Sci.}~{\bf 52},  2045--2051 (2005).

\bibitem{Watanabe09}
Watanabe, S., Ishikawa, S., Aono, H., Takeda, S., Odaka, H., Kokubun, M.,
  Takahashi, T., Nakazawa, K., Tajima, H., Onishi, M., and Kuroda, Y., ``High
  energy resolution hard {X-ray} and gamma-ray imagers using {CdTe} diode
  devices,'' {\em IEEE Trans. Nucl. Sci.}~{\bf 56},  777--782 (2009).

\bibitem{Takeda09}
Takeda, S., Aono, H., Okuyama, S., n.~Ishikawa, S., Odaka, H., Watanabe, S.,
  Kokubun, M., Takahashi, T., Nakazawa, K., Tajima, H., and Kawachi, N.,
  ``Experimental results of the gamma-ray imaging capability with a {Si/CdTe}
  semiconductor {Compton} camera,'' {\em IEEE Trans. Nucl. Sci.}~{\bf 56},
  783--790 (2009).

\bibitem{HXD}
Kamae, T., Ezawa, H., Fukazawa, Y., M, H., Idesawa, E., Iyomoto, N., et~al.,
  ``{Astro-E} hard {X}-ray detector,'' {\em Proc. SPIE}~{\bf 2806},  314
  (1996).

\bibitem{VA94}
Toker, O., Masciocchi, S., Nyg{\aa}rd, E., Rudge, A., and Weilhammer, P.,
  ``{VIKING}, a {CMOS} low noise monolithic 128 channel frontend for {Si}-strip
  detector readout,'' {\em Nucl. Instrum. Methods A}~{\bf 340},  572--579
  (1994).

\bibitem{Tajima04}
Tajima, H., Nakamoto, T., Tanaka, T., Uno, S., Mitani, T., do~Couto~e Silva,
  E., et~al., ``Performance of a low noise front-end {ASIC} for {Si/CdTe}
  detectors in {Compton} gamma-ray telescope,'' {\em IEEE Trans. Nucl.
  Sci.}~{\bf 51},  842--847 (2004).

\bibitem{Armstrong73}
Armstrong, T.~W. et~al. {\em J. Geophys. Res.}~{\bf 78},  2715 (1973).

\bibitem{Fukazawa09}
Fukazawa, Y. et~al., ``Modeling and reproducibility of {Suzaku} {HXD PIN/GSO}
  background,'' {\em Pub. of Astro. Soc. of Japan}~{\bf 61},  S17 (2009).

\bibitem{Murakami03}
Murakami, M., Kobayashi, Y., Kokubun, M., Takahashi, I., Okada, Y., Kawaharada,
  M., Nakazawa, K., Watanabe, S., Sato, G., Kouda, M., Mitani, T., Takahashi,
  T., Suzuki, M., Tashiro, M., Kawasoe, S., Nomachi, M., and Makishima, K.,
  ``Activation properties of schottky {CdTe} diodes irradiated by 150 {MeV}
  protons,'' {\em IEEE Trans. Nucl. Sci.}~{\bf 50},  1013--1019 (2003).

\bibitem{INTEGRAL}
{Winkler}, C., {Courvoisier}, T.~J.-L., {Di Cocco}, G., {Gehrels}, N.,
  {Gim{\'e}nez}, A., {Grebenev}, S., {Hermsen}, W., {Mas-Hesse}, J.~M.,
  {Lebrun}, F., {Lund}, N., {Palumbo}, G.~G.~C., {Paul}, J., {Roques}, J.-P.,
  {Schnopper}, H., {Sch{\" o}nfelder}, V., {Sunyaev}, R., {Teegarden}, B.,
  {Ubertini}, P., {Vedrenne}, G., , and {Dean}, A.~J., ``The {INTEGRAL}
  mission,'' {\em Astronomy and Astrophysics}~{\bf 411},  L1--L6 (2003).

\end{thebibliography}
\bibliographystyle{spiebib}   

\end{document}